\documentclass[preprint2]{proto}
\usepackage{times}
\newcommand{\refs}{\par\noindent\hangindent=1pc\hangafter=1}
\newcommand{\gcc}{\rm g\:cm^{-2}}

\newcommand{\cth}{c_{\rm th}}

\newcommand{\tff}{t_{\rm ff}}

\newcommand{\beq}{\begin{equation}}
\newcommand{\eeq}{\end{equation}}
\newcommand{\beqa}{\begin{eqnarray}}
\newcommand{\eeqa}{\end{eqnarray}}
\newcommand{\e}{$^{-1}$}
\newcommand{\ee}{$^{-2}$}
\newcommand{\eee}{$^{-3}$}

\newcommand{\krho}{{k_\rho}}
\newcommand{\phig}{\phi_{\rm geom}}

\newcommand{\alv}{\alpha_{\rm vir}}
\newcommand{\alvcl}{\alpha_{\rm vir,cl}}

\newcommand{\mbe}{M_{\rm BE}}

\newcommand{\mdsd}{\dot m_{*d}}
\newcommand{\scl}{\Sigma_{\rm cl}}

\newcommand{\rcl}{R_{\rm cl}}

\newcommand{\ecore}{\epsilon_c}

\newcommand{\AMM}{$\rm NH_3$}
\newcommand{\NNH}{$\rm N_2H^+$}
\newcommand{\NND}{$\rm N_2D^+$}
\newcommand{\Dfrac}{$D_{\rm frac}$}
\newcommand{\kms}{$\rm km\:s^{-1}$}

\newcommand{\COII}{\mbox{C$^{18}$O}}

\newcommand{\mcr}{M_{\rm cr}}
\newcommand{\mubcr}{\bar\mu_{\rm cr}}

\newcommand{\pc}{{\rm pc}}
\newcommand{\rpc}{R_{\rm pc}}
\newcommand{\sigcl}{\sigma_{\rm cl}}
\newcommand{\spc}{\sigma_{\rm pc}}
\newcommand{\va}{v_{\rm A}}

\newcommand{\mdch}{\dot m_{\rm ff}}

\begin{document}

\title{\textbf{\LARGE Massive Star Formation}}

\author {\textbf{\large Jonathan C. Tan$^1$, Maria T. Beltr\'an$^2$, Paola Caselli$^3$, Francesco Fontani$^2$, Asunci\'on Fuente$^4$, Mark R. Krumholz$^5$, Christopher F. McKee$^6$, Andrea Stolte$^7$}}
\affil{\small\em $^1$University of Florida; $^2$INAF-Osservatorio Astrofisico di Arcetri; $^3$University of Leeds; $^4$Observatorio Astron\'omico Nacional; $^5$University of California, Santa Cruz; $^6$University of California, Berkeley; $^7$Argelander Institut f\"ur Astronomie, Universit\"at Bonn}

\begin{abstract}
\baselineskip = 11pt
\leftskip = 0.65in 
\rightskip = 0.65in \parindent=1pc {\small 
The enormous radiative and mechanical luminosities of massive stars
impact a vast range of scales and processes, from the reionization of
the universe, to the evolution of galaxies, to the regulation of the
interstellar medium, to the formation of star clusters, and even to
the formation of planets around stars in such clusters. Two main
classes of massive star formation theory are under active study, {\it
  Core Accretion} and {\it Competitive Accretion}. In Core Accretion,
the initial conditions are self-gravitating, centrally concentrated
cores that condense with a range of masses from the surrounding,
fragmenting clump environment. They then undergo relatively ordered
collapse via a central disk to form a single star or a small-$N$
multiple. In this case, the pre-stellar core mass function has a
similar form to the stellar initial mass function. In Competitive
Accretion, the material that forms a massive star is drawn more
chaotically from a wider region of the clump without passing through a
phase of being in a massive, coherent core. In this case, massive star
formation must proceed hand in hand with star cluster formation. If
stellar densities become very high near the cluster center, then
collisions between stars may also help to form the most massive
stars. We review recent theoretical and observational progress towards
understanding massive star formation, considering 
physical and chemical processes, comparisons with low and
intermediate-mass stars, and connections to star cluster
formation.\\~\\~\\}
\end{abstract}

\section{\textbf{INTRODUCTION}}

Across the universe, massive stars play dominant roles in terms of
their feedback and their synthesis and dispersal of heavy elements.
Achieving a full theoretical understanding of massive star formation
is thus an important goal of contemporary astrophysics. This effort
can also be viewed as a major component of the development of a
general theory of star formation that seeks to explain the birth of
stars of all masses and from all varieties of star-forming
environments.

Two main classes of theory are under active study, {\it Core
  Accretion} and {\it Competitive Accretion}. In Core Accretion,
extending ``standard'' low-mass star formation theory ({\em Shu et
  al.}, 1987), the initial conditions are self-gravitating, centrally
concentrated {\it cores} of gas that condense with a range of masses
from a fragmenting {\it clump} (i.e., protocluster) environment. These
cores then undergo gravitational collapse via a central disk, to form
a single star or small-$N$ multiple. The pre-stellar core (PSC) mass
function (CMF) has a shape similar to the stellar initial mass
function (IMF). In Competitive Accretion, gas that forms a massive
star is drawn chaotically from a wider region of the clump, without
ever being in a massive, coherent, gravitationally bound, starless
core. Also, a forming massive star is always surrounded by a swarm of
low-mass protostars. Competitive Accretion is sometimes said to lead
naturally to the IMF ({\it Bonnell et al.,} 2001; 2007): then the
total mass of massive stars must be a small fraction of the total
stellar mass formed from the clump. If the density of protostars
congregating near the cluster center becomes sufficiently high, then
stellar collisions may also assist in forming the most massive stars.

Recent advances in theoretical/numerical modeling of massive star
formation involve inclusion of more physical processes, like radiation
pressure, magnetic fields and protostellar outflows.  Observationally,
progress has resulted from telescopes such as {\it Spitzer}, {\it
  Herschel}, {\it SOFIA}, {\it ALMA} and the {\it VLA}.  Galactic
plane surveys have yielded large samples of candidate massive
protostars and their birth clouds.

This review aims to summarize massive star formation research,
focusing on developments since the reviews of {\em Beuther et al.}
(2007), {\em Zinnecker and Yorke} (2007) and {\em McKee \& Ostriker}
(2007). We do not discuss formation of the first stars, which are
thought to have been massive (e.g., {\it Bromm}, 2013). Given the
complexity of massive star formation, detailed comparison of
theoretical predictions with observational results is needed for
progress in understanding which accretion mechanism(s) is relevant and
which physical and chemical processes are important. We thus first
overview basic observed properties of massive star-forming regions
(\S\ref{S:environ}), which set boundary conditions on theoretical
models. Next we present a theoretical overview of physical processes
likely involved in forming massive stars (\S\ref{S:theory}), including
the different accretion models, protostellar evolution and feedback,
and results from numerical simulations. We then focus on observational
results on the earlier, i.e., initial condition (\S\ref{S:initial})
and later, i.e., accretion (\S\ref{S:accretion}) stages of massive
star formation. Here we discuss astrochemical modeling, as well as
general comparisons of massive star formation with
intermediate/low-mass star formation.  The relation of massive star
formation to star cluster formation is examined in
\S\ref{S:cluster}. We conclude in \S\ref{S:conclusions}.

\subsection{The Birth Environments of Massive Stars}\label{S:environ}

\begin{figure*}
\epsscale{2.05}
\plotone{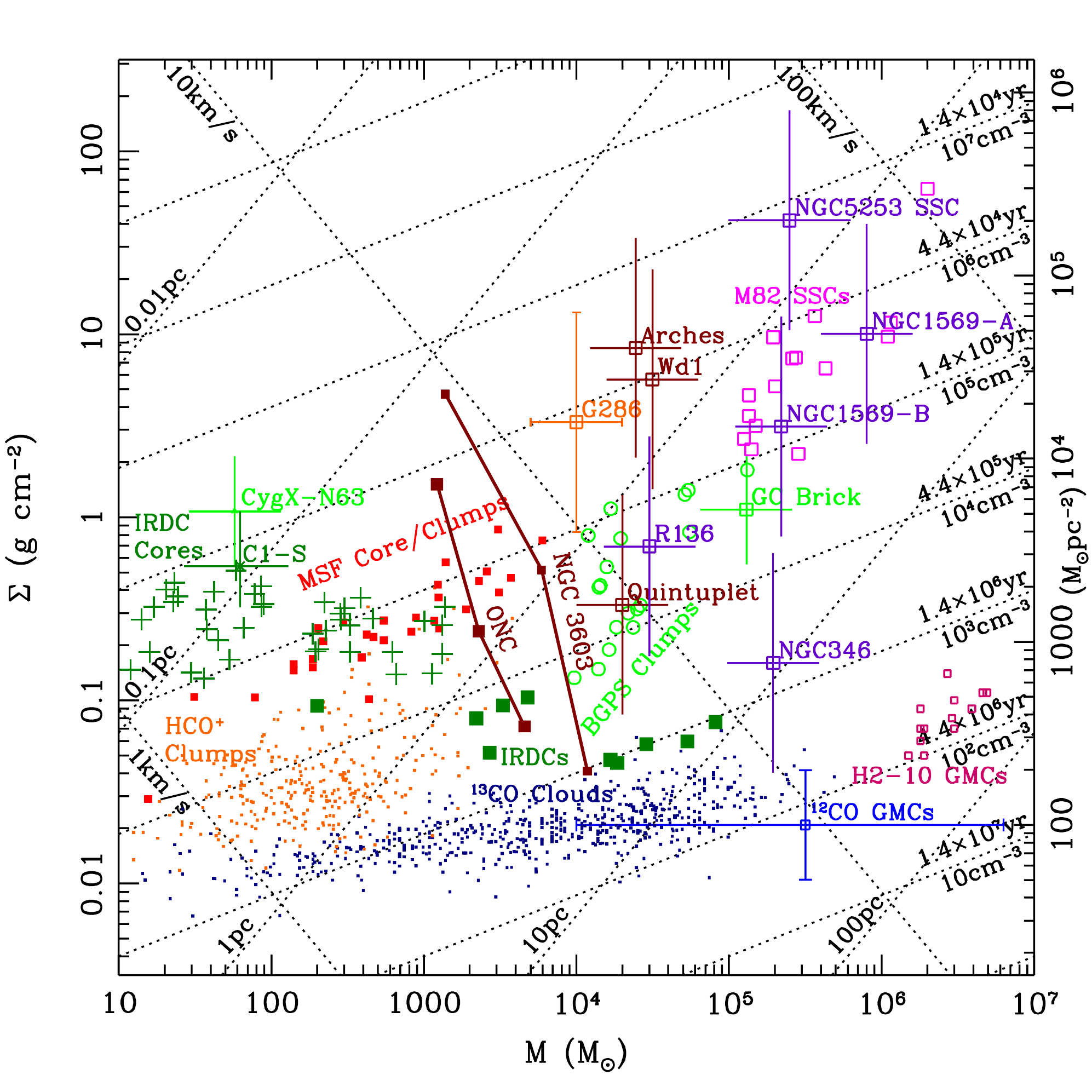}
\caption{\small \label{fig:overview} 
The Environments of Massive Star Formation.  Mass surface density,
$\Sigma\equiv M/(\pi R^2)$, is plotted versus mass, $M$. Dotted lines
of constant radius, $R$, H number density, $n_{\rm H}$ (or free-fall
time, $t_{\rm ff}= (3\pi/[32G\rho])^{1/2}$), and escape speed, $v_{\rm
  esc} = (10/\alpha_{\rm vir})^{1/2}\sigma$, are shown. Stars form
from molecular gas, which in the Galaxy is mostly organized into
GMCs. Typical $\rm ^{12}CO$-defined GMCs have $\Sigma\sim
100\:M_\odot\:{\rm pc^{-2}}$ ({\it Solomon et al.,} 1987) (see {\it
  Tan et al.,} 2013a for detailed discussion of the methods for
estimating $\Sigma$ for the objects plotted here), although denser
examples have been found in Henize 2-10 ({\it Santangelo et al.,}
2009). The $\rm ^{13}CO$-defined clouds of {\it Roman-Duval et al.}
(2010) are indicated, along with $\rm HCO^+$ clumps of {\it Barnes et
  al.,} (2011), including G286.21+0.17 ({\it Barnes et al.,}
2010). Along with G286, the BGPS clumps ({\it Ginsburg et al.,} 2012)
and the Galactic Center ``Brick'' ({\it Longmore et al.,} 2012) are
some of the most massive, high-$\Sigma$ gas clumps known in the Milky
Way. Ten example Infrared Dark Clouds (IRDCs) ({\it Kainulainen and
  Tan,} 2013) and their internal core/clumps ({\it Butler and Tan,}
2012) are shown, including the massive, monolithic, highly-deuterated
core C1-S ({\it Tan et al.,} 2013b). 
CygX-N63, a core with similar mass and size as C1-S, appears to be
forming a single massive protostar ({\it Bontemps et al.,} 2010;
{\it Duarte-Cabral et al.,} 2013).
The IRDC core/clumps overlap with Massive Star-Forming (MSF)
core/clumps ({\it Mueller et al.,} 2002).  Clumps may give rise to
young star clusters, like the ONC (e.g., {\it Da Rio et al.,} 2012)
and NGC 3603 ({\it Pang et al.,} 2013) (radial structure is shown from
core to half-mass, $R_{1/2}$, to outer radius), or even more massive
examples, e.g., Westerlund 1 ({\it Lim et al.,} 2013), Arches ({\it
  Habibi et al.,} 2013), Quintuplet ({\it Hu{\ss}mann et al.,}~2012)
(shown at $R_{1/2}$), that are in the regime of ``super star
clusters'' (SSCs), i.e., with $M_*\gtrsim 10^4\:M_\odot$. Example SSCs
in the Large Magellanic Cloud (LMC) (R136, {\it Andersen et al.,}
2009) and Small Magellanic Cloud (SMC) (NGC 346, {\it Sabbi et al.,}
2008) display a wide range of $\Sigma$, but no evidence of IMF
variation (\S\ref{S:IMF}).  Even more massive clusters can be found in
some dwarf irregular galaxies, such as NGC 1569 ({\it Larsen et al.,}
2008) and NGC 5253 ({\it Turner and Beck,} 2004), and starburst galaxy
M82 ({\it McCrady and Graham,} 2007).  }
\end{figure*}

The basic physical properties of regions observed to be forming or
have formed massive stars, i.e., gas clumps and young star clusters,
are shown in Fig.~\ref{fig:overview}, plotting mass surface density,
$\Sigma = M/(\pi R^2)$, of the structure versus its mass, $M$.  Stars,
including massive stars, form in molecular gas, that is mostly found
in giant molecular clouds (GMCs) with $\Sigma\sim 0.02\:\gcc$. Note
$1.0\:\gcc \equiv 4790\:M_\odot\:{\rm pc^{-2}}$, for which $N_{\rm H}
= 4.27\times 10^{23}\:{\rm cm^{-2}}$ (assuming $n_{\rm He}=0.1 n_{\rm
  H}$ so mass per H is $\mu_{\rm H}=2.34\times 10^{-24}\:{\rm g}$) and
visual extinction is $A_V=(N_{\rm H}/2.0\times 10^{21}$~cm$^{-2})$~mag
$= 214$~mag. However, star formation is seen to be localized within
star-forming clumps within GMCs, which typically have $\Sigma_{\rm
  cl}\sim 0.1 - 1\:\gcc$. Some massive systems, usually already-formed
star clusters, have $\Sigma$ up to $\sim 30\:\gcc$.

In terms of $\Sigma_{\rm cl}$ (in $\rm g\; cm^{-2}$) and $M_{\rm
  cl,3}=M_{\rm cl}/(1000\:M_\odot)$, the radius and (H number) density
of a spherical clump are
\beqa
R_{\rm cl}&=&0.258 M_{\rm cl,3}^{1/2} \Sigma_{\rm cl}^{-1/2}~~\pc,\label{eq:rcl}\\
\bar{n}_{\rm H,cl}&=&4.03\times 10^5 \Sigma_{\rm cl}^{3/2} M_{\rm cl,3}^{-1/2}~~~\mbox{cm\eee}.
\eeqa 
Gas clumps massive enough to form a cluster of mass $M_{\rm *cl}\sim
500\:M_\odot$, i.e., with median expected maximum stellar mass
$\sim30\:M_\odot$
(for Salpeter IMF from 0.1 to 120~$M_\odot$), are thus $\sim$0.3~pc in
size (if $\Sigma_{\rm cl}\sim 1\:{\rm g\:cm^{-2}}$ and efficiency
$\epsilon_{\rm *cl}\equiv M_{\rm *cl}/M_{\rm cl}\sim 0.5$), only
moderately larger than the $\sim$0.1~pc sizes of well-studied low-mass
starless cores in regions such as Taurus ({\it Bergin and Tafalla,}
2007). However, mean densities in such clumps are at least ten times
larger.

\section{\textbf{THEORETICAL OVERVIEW}}\label{S:theory}

\subsection{Physical Processes in Self-Gravitating Gas}

The importance of self-gravity in a cloud of mass $M$ and
radius $R$ can be gauged by the virial parameter
\beq
\alv\equiv 5\sigma^2 R/(GM)= 2 a E_K/E_G,
\label{eq:alv}
\eeq 
where $\sigma$ is 1D mass-averaged velocity dispersion, $a\equiv
E_G/(3GM^2/[5R])$ is the ratio of gravitational energy, $E_G$ (assuming
negligible external tides), to that of a uniform sphere, and $E_K$ is
the kinetic energy ({\it Bertoldi and McKee,} 1992). Often $\alv$ is
set as $2E_K/E_G$, with the advantage of clearly denoting bound
($E_K<E_G$) and virialized ($E_K=\frac 12 E_G$) clouds, but the
disadvantage that $a$ is difficult to observe. For spherical clouds
with a power-law density distribution, $\rho\propto r^{-\krho}$, $a$
rises from 1 to $\frac 53$ as $\krho$ goes from 0 (uniform density) to
2 (singular isothermal sphere). A cloud in free-fall has
$\alv\rightarrow 2a$ from below as time progresses. The cloud's escape
velocity is $v_{\rm esc}=(10/\alv)^{1/2}\sigma$.

The velocity dispersion in a cloud is thus given by
\beq
\sigma\equiv (\pi \alv G \Sigma R/5)^{1/2},
\label{eq:sigma}
\eeq 
where we have used the identity symbol to emphasize that this follows
from the definition of $\alv$. Clouds that are gravitationally bound
with $\alv\sim 1$ and that have similar surface densities, then naturally
satisfy a line-width size (LWS) relation $\sigma\propto R^{1/2}$ ({\it
  Larson,} 1985), consistent with observations ({\it McKee and
  Ostriker,} 2007; {\it Heyer et al.,} 2009). {\it McKee et al.}
(2010) termed this the virialized LWS relation for $\alv=1$, as
suggested by {\it Heyer et al.} (2009). By contrast, the standard
turbulence-dominated LWS relation, $\sigma=\spc \rpc^{1/2}$, where
$\spc\simeq 0.72$~km s\e\ in the Galaxy ({\it McKee and Ostriker,}
2007), is independent of $\Sigma$. The virialized LWS relation applies
for mass surface densities $\Sigma>\Sigma_{\rm LWS}=(5/[\pi G])(\spc^2/1\;\pc)\simeq 0.040$~g~cm\ee.  Since regions of massive star
formation have column densities substantially greater than this, they
lie above the turbulence-dominated LWS relation, as observed ({\it
  Plume et al.,} 1997).

Evaluating eq.~(\ref{eq:sigma}) for massive star-forming regions,
%
\beq
\sigma=1.826\alv^{1/2}(M_3\Sigma)^{1/4}~~~\mbox{km s\e},
\eeq
yields supersonic motions, as the isothermal sound speed
$\cth=0.188(T/10\;\mbox{K})^{1/2}$~km~s\e\ and $T\la 30$~K for gas 
not too close to a massive star. These motions cannot be primarily
infall, since clump infall (\S\ref{S:infall}) and star-formation rates per free-fall time are quite
small
({\it Krumholz and Tan,} 2007).
It is therefore likely that regions of
massive star formation are dominated by supersonic turbulence. If so,
then the gas inside a clump of radius $\rcl$ will obey a LWS relation
$\sigma=(R/\rcl)^{q}\sigcl$ with $q\leq \frac 12$ ({\it Matzner,}
2007), so
\beq
\alv=(5/\pi) (\sigcl^2/[\rcl G\Sigma])= (\scl/\Sigma)\alvcl.
\eeq 
At a typical point in the clump, the density is less than average, so
$\alv$ of a sub-region of size $R$ is
$>(\rcl/R)\alvcl$. Even if the clump is bound, the sub-region is not.
However,
for $q= \frac 12$, a sub-region compressed so
that $\Sigma\ga\scl$ has $\alv\la \alvcl$ and is bound if the
clump is; for $q< \frac 12$, extra
compression is needed to make a bound sub-region.

Isothermal clouds more massive than the critical mass, $\mcr$, cannot
be in hydrostatic equilibrium and will collapse.  In this case, $\mcr$
is termed the Bonnor-Ebert mass, given by
\beq 
\mcr=\mbe= \mubcr \cth^3 (G^3\bar\rho)^{-1/2} = \mubcr \cth^4 (G^3\bar P)^{-1/2},
\eeq 
where $\bar P$ is mean pressure in the cloud, and $\bar\mu_{\rm
  cr}=1.856$ ({\it McKee and Holliman,} 1999; $\mcr$ can also be
expressed in terms of density at the cloud's surface: then 
$\bar\mu_{\rm cr}=1.182$).  For non-magnetic clouds this
relation can be generalized to an arbitrary equation of state by
replacing $\cth^2$ with $\sigma^2$. One can show that $\mubcr$
corresponds to a critical value of the virial parameter, $\alpha_{\rm
  vir,\,cr}=5(3/[4\pi])^{1/3}\mubcr^{-2/3}$; clouds with
$\alv<\alpha_{\rm vir,\,cr}$ will collapse.  
For example, 
equilibrium
isothermal clouds have $\alpha>\alpha_{\rm vir,\,cr}=2.054$.

The critical mass associated with magnetic fields can be expressed in
two ways (e.g., {\it McKee and Ostriker,} 2007): as $M_\Phi=\Phi/(2\pi
G^{1/2})$, where $\Phi$ is the magnetic flux, or as
$M_B=M_\Phi^3/M^2$, which can be rewritten as
\beq
M_B=(9/[16\surd\pi])\; \va^3 (G^3\bar\rho)^{-1/2}, 
\eeq
where $\va$ is the Alfv\'en velocity.  Magnetically critical clouds
have $M=M_\Phi=M_B$.  Most regions with Zeeman observations are
magnetically supercritical ({\it Crutcher,} 2012); i.e., $M>M_\Phi$ so
that the field, on its own, cannot prevent collapse. Magnetized
isothermal clouds have $\mcr\simeq \mbe+M_\Phi$, and since most GMCs
are gravitationally bound (e.g., {\it Roman-Duval et al.,} 2010; {\it
  Tan et al.,} 2013a), magnetized and turbulent, they are expected to
have $M\simeq\mcr\simeq 2 M_\Phi$ ({\it McKee,} 1989). Magnetically
subcritical clouds can evolve to being supercritical by flows along
field lines and/or by ambipolar diffusion. In a quiescent medium, the
ambipolar diffusion time is about 10 free-fall times ({\it
  Mouschovias,} 1987); this time scale is reduced in the presence of
turbulence (e.g., {\it Fatuzzo and Adams,} 2002; {\it Li et al.,}
2012b).  Lazarian and collaborators (e.g., {\it de Gouveia Dal Pino et
  al.,} 2012 and references therein) have suggested super-Alfv\'enic
turbulence drives rapid reconnection that can efficiently remove
magnetic flux from a cloud.

Self-gravitating clouds in virial equilibrium have a mean total
pressure (thermal, turbulent and magnetic) that is related to the
total $\Sigma$ via ({\it McKee and Tan,} 2003 [MT03]), 
\beq
\bar{P}\equiv (3\pi/20) f_g \phig \phi_B \alv G\Sigma^2,
\eeq 
where $f_g$ is the fraction of total mass surface density in gas (as
opposed to stars), $\phig$ is an order unity numerical factor that
accounts for the effect of nonspherical geometry, $\phi_B \simeq
1.3+3/(2 {\cal M}_{A}^2)$ accounts for the effect of magnetic fields,
and ${\cal M}_{A}=\surd 3(\sigma/\va)$ is the Alfv\'en Mach number.  A
magnetized cloud with the same total pressure and surface density as a
non-magnetic cloud will therefore have a virial parameter that is
smaller by a factor $\phi_B$. Clouds that are observed to have small
$\alv$ (e.g., {\it Pillai et al.,} 2011; see also
\S\ref{S:physical_starless}) are therefore either in the very early
stages of gravitational collapse or are strongly magnetized.

The characteristic time for gravitational collapse is the free-fall
time. For a spherical cloud, this is
\beq
\bar\tff=(3\pi/[32G\bar\rho])^{1/2}=6.85\times 10^4 M_3^{1/4} \Sigma^{-3/4}\:{\rm yr},
\eeq
where the bar on $\tff$ indicates that it is given in terms of the
mean density of the mass $M$.  The free-fall velocity is
\beq
v_{\rm ff} = (2GM/r)^{1/2}
=5.77(M_3\Sigma)^{1/4}~~~\mbox{km s\e}.
\eeq
An isothermal filament with mass/length
$m_\ell>2\cth^2/G=16.4(T/10\mbox{ K})$ $M_\odot$ ~pc\e\ cannot be in
equilibrium and will collapse. Its free-fall time and velocity are
$(1/2) (G\bar\rho)^{-1/2}$ and $2 [G m_\ell \ln (r_0/r)]^{1/2} =
1.3 [ m_{\ell,100} \ln(r_0/r)]^{1/2}\:\mbox{km s\e}$, with $r_0$ the
initial radius of collapsing gas and
$m_{\ell,100}=m_\ell/100\,M_\odot\mbox{pc\e}$. Infall velocities much
less than this indicate either collapse has just begun or that it is
quasi-static.

\subsection{Formation Mechanisms}

A key parameter in both Core and Competitive Accretion is the characteristic
accretion rate in a cloud with $M\geq\mcr$,
\beqa
\mdch&=& M/\bar\tff = (8G/\surd\pi)^{1/2}(M\Sigma)^{3/4},\label{eq:mdch1}\\ 
&=&1.46\times 10^{-2}(M_3\Sigma)^{3/4}~~~M_\odot\mbox{ yr\e}.\label{eq:mdch2}
\eeqa 
In Competitive Accretion models, the star-forming clump undergoes
global, typically free-fall, collapse, so this is the characteristic
accretion rate in the entire forming cluster. In Core Accretion
models, this is the characteristic accretion rate to the central star
and disk in the core, with the accreted gas then supplied to just one
or a few protostars.  The properties of the surrounding clump are
assumed to be approximately constant during the formation of the star.

The corresponding accretion time, $t_{\rm acc}\propto M/\dot{m}_{\rm
  ff}\propto \bar\tff \propto M^{1/4}$, is a weak function of mass for
clouds of a given $\Sigma$.  Note, a singular isothermal sphere has
$\rho\propto r^{-2}$ so its collapse leads to
$\mdch\propto(M\Sigma)^{3/4}=$ const ({\it Shu,} 1977).

\vspace{-0.01in}
\subsubsection{Core Accretion}

The principal assumption of Core Accretion models is that the initial
conditions for intermediate and massive star formation are
gravitationally bound cores, scaled up in mass from the low-mass
examples known to form low-mass stars. Different versions of these
models invoke varying properties of the cores, including their
expected densities, density profiles, sources of internal pressure and
dynamical states.  A distinguishing feature of these models is that
the pre-stellar CMF is hypothesized to be similar in shape to the
stellar IMF, with stellar masses being $m_*=\ecore M_{c}$, where
$\ecore\sim 0.5$, perhaps set by protostellar outflow feedback ({\it
  Matzner and McKee,} 2000; see \S\ref{S:feedback}). This feature of
some kind of one-to-one correspondence between the CMF and IMF is an
underlying assumption of recent theories of the IMF, which predict the
CMF based on the conditions needed to form bound cores in a turbulent medium
(e.g., {\it Padoan and Nordlund,} 2007; cf., {\it Clark et al.,} 2007).

There are at least two main differences between low and high-mass star
formation: First, for sufficiently massive stars, the Kelvin-Helmholtz
time can be less than the accretion time, so the star accretes while
on the main sequence ({\it Kahn,} 1974). Second, cores forming massive
stars are large enough that internal turbulence can dominate thermal
motions ({\it Myers and Fuller,} 1992; {\it Caselli and Myers,} 1995).
Extending the work of these authors, {\it McKee and Tan} (2002; {\it
  MT03}) developed the Turbulent Core model, based on the assumptions
that the internal pressure is mostly nonthermal, in the form of
turbulence and/or magnetic fields, and that the 
initial core
is reasonably close to internal virial equilibrium, so that its
structure can be approximated as a singular polytropic sphere.  Also,
approximate pressure equilibrium with the surrounding clump is
assumed, which thus normalizes the size, density and velocity
dispersion of a core of given mass to $\Sigma_{\rm cl}$.
{\it MT03} focused on the case in which 
$\rho \propto r^{-\krho}$ with $\krho=1.5$, similar to observed values
(\S\ref{S:physical_starless}); for this case,
$\Sigma_{c}=1.22\Sigma_{\rm cl}$.  
For example, the core radius, 
given by eq. (\ref{eq:rcl}) with core properties 
in place of those of the clump, can be expressed in terms of core mass and
clump surface density: $R_{c}=0.074(M_{\rm c,\,2}/\Sigma_{\rm
  cl})^{1/2}$ pc.

The characteristic accretion rate in Core Accretion models is given by
eq.~(\ref{eq:mdch1}). In the {\it Shu} (1977) model, based on collapse
of a singular isothermal sphere, the actual accretion rate is
$0.38\mdch$. This result ignores the contraction needed to create the
sphere. {\it Tan and McKee} (2004) argued (in the context of
primordial star formation, but similar reasoning may apply locally)
that it was more reasonable to include the formation phase of the
collapsing cloud using one of {\it Hunter's} (1977) subsonic collapse
solutions, which has an accretion rate 2.6 times larger and gives an
accretion rate onto the star + disk system of $\mdsd\simeq\mdch$. For
collapse that begins from a marginally stable Bonnor-Ebert sphere,
$\mdsd$ is initially $\gg \mdch$, but then falls to about the Shu
rate. For the Turbulent Core model, the dependence of the accretion
rate on $M\Sigma$ can be re-expressed in terms of the current value of
the idealized collapsed-mass that has been supplied to the central
disk in the zero-feedback limit, $M_{*d}$, (note, the actual protostar
plus disk mass accretion rate is $\dot{m}_{*d} = \epsilon_{*d}
\dot{M}_{*d}$ and the integrated protostar plus disk mass is $m_{*d} =
\bar{\epsilon}_{*d} M_{*d}$) and $\Sigma_{\rm cl}$. For $\krho=1.5$
and allowing for the effects of magnetic fields ({\it MT03}), this
gives
\beq \mdsd = 1.37\times 10^{-3} \epsilon_{*d} (M_{c,2} \Sigma_{\rm cl})^{3/4} (M_{*d}/M_c)^{1/2}\:M_\odot\:{\rm yr}^{-1};
\label{eq:mdsd}
\eeq 
for $\epsilon_{*d}=1$, this corresponds to $\mdsd=0.64\mdch$.  
If the disk mass is assumed to be a constant fraction, $f_d$,
of the stellar mass, then the actual accretion rate to the protostar
is $\dot{m}_* = (1/[1+f_d])\dot{m}_{*d}$. A value of $f_d\simeq 1/3$,
i.e., a relatively massive disk, is expected in models where angular
momentum transport is due to moderately self-gravitating disk
turbulence and larger-scale spiral density waves.

Two challenges faced by Core Accretion are: (1) What
prevents a massive core, perhaps containing $\sim 10^2$ Jeans masses,
fragmenting into a cluster of smaller stars?  This will be
addressed in \S\ref{S:feedback}. (2) Where are the accretion disks
expected around forming single and binary massive stars? 
Disks have been discovered around some massive stars, but it has not
been shown that they are ubiquitous (\S\ref{S:disks}).

\subsubsection{Competitive Accretion}\label{S:competitive}

Competitive Accretion ({\it Bonnell et al.,} 2001) involves protostars
accreting ambient clump gas at a rate
\beq 
\dot{m}_{*d}=\pi \rho_{\rm cl} v_{\rm rel} r_{\rm acc}^2, 
\eeq 
where $v_{\rm rel}$ is the relative velocity of stars with respect to
clump gas, $\rho_{\rm cl}$ is the local density, and $r_{\rm acc}$ is
the accretion radius. Two limits for $r_{\rm acc}$ were proposed: (1)
Gas-dominated regime (set by tidal radius): $r_{\rm acc}\simeq r_{\rm
  tidal} = 0.5[m_*/M_{\rm cl}(R)]^{1/3}R$, where $R$ is the distance
of the star from the clump center; (2) star-dominated regime (set by
Bondi-Hoyle accretion radius): $r_{\rm acc}\simeq r_{\rm BH} =
2Gm_*/(v_{\rm rel}^2+c_s^2)$. The star-dominated regime was suggested
to be relevant for massive star formation---the stars destined to
become massive being those that tend to settle to protocluster
centers, where high ambient gas densities are maintained by global
clump infall. The accretion is assumed to be terminated by stellar
feedback or by fragmentation induced starvation ({\it Peters et al.,}
2010b).

In addition to forming massive stars, {\it Bonnell et al.}
(2001; 2007) proposed Competitive Accretion is also responsible for
building up the IMF for $m_*\gtrsim \mbe$. These
studies have since been developed to incorporate additional physics
(see \S\ref{S:numerical}) and include comparisons to both the IMF and
binary properties of the stellar systems ({\it Bate,} 2012).

{\it Bonnell et al.} (2004) tracked the gas that joined
the massive stars in their simulation, showing it was initially
widely distributed throughout the clump, so the final mass of the
star did not depend on the initial core mass present when it first
started forming. Studies of the gas cores seen in simulations
exhibiting Competitive Accretion have been carried out by {\it Smith
  et al.} (2011, 2013), with non-spherical, filamentary morphologies
being prevalent, along with total accretion being dominated by that
accreted later from beyond the original core volume. Other predictions
of the Competitive Accretion scenario are relatively small accretion
disks, with chaotically varying orientations, which would also be
reflected in protostellar outflow directions. Massive stars would
always be observed to form at the center of a cluster in which the
stellar mass was dominated by low-mass stars.

As Competitive Accretion is ``clump-fed'', we express the 
average
accretion rate of a star of final mass $m_{*f}$ via
\beqa \langle \dot{m}_{*d} \rangle & = &
\epsilon_{\rm ff} \mdch m_{*f} /(\epsilon_{\rm cl} M_{\rm
  cl})\label{eq:mdot_comp2}\\ &\rightarrow & 1.46 \times 10^{-4}
\epsilon_{\rm ff,0.1} \frac{m_{*f,50}}{\epsilon_{\rm cl,0.5}}
\frac{\Sigma_{\rm cl}^{3/4}}{M_{\rm cl,3}^{1/4}}\:M_\odot\:{\rm
  yr}^{-1}\nonumber \eeqa
(see also {\it Wang et al.,} 2010), where $\epsilon_{\rm ff}$ is the star formation efficiency per
free-fall time and $\epsilon_{\rm cl}$ is the final star formation
efficiency from the clump. {\it Krumholz and Tan} (2007) estimated
$\epsilon_{\rm ff}\simeq 0.04$ in the ONC.  The average accretion rate
($4.6\times 10^{-5}\:M_\odot\:{\rm yr}^{-1}$) of the most massive star
(46.4~$M_\odot$) in the {\it Wang et al.} (2010) simulation with
outflow feedback ($\Sigma_{\rm cl}=0.08\:{\rm g\:cm^{-2}}$, $M_{\rm
  cl}=1220\:M_\odot$, $\epsilon_{\rm cl}=0.18$, $\epsilon_{\rm
  ff}=0.08$) agrees with 
eq.~(\ref{eq:mdot_comp2})
to within 10\%.  
This
shows that a major
difference of the Competitive Accretion model of massive star
formation from Core Accretion is that its average accretion rate to
the star is much smaller (cf., eq.~\ref{eq:mdsd}).

This low rate of competitive accretion was noted before in the context
of accretion from a turbulent medium with $\alv\sim 1$, as is observed
in most star-forming regions ({\it Krumholz et al.,} 2005a).  {\it
  Bonnell et al.} (2001) came to essentially the same conclusion by
noting competitive accretion would not be fast enough to form massive
stars, if stars were virialized in the cluster potential (i.e., high
$v_{\rm rel}$). They suggested efficient star formation
($\epsilon_{\rm ff}\sim 1$) occurs in regions of global gravitational
collapse with negligible random motions.  {\it Wang et al.} (2010), by
including the effects of protostellar outflows and moderately strong
magnetic fields that slowed down star cluster formation, found massive
star formation via competitive accretion occurred relatively slowly
over about 1~Myr (eq.~\ref{eq:mdot_comp2}). Accretion to the clump
center was fed by a network of dense filaments, even while the overall
clump structure remained in quasi virial equilibrium.
As discussed further in \S\ref{S:numerical}, these results may depend
on the choice of initial conditions, such as the degree of
magnetization and/or use of an initially smooth density field, which
minimizes the role of turbulence.

Another challenge for Competitive Accretion is the effect of
feedback. {\it Edgar and Clarke} (2004) noted radiation pressure
disrupts dusty Bondi-Hoyle accretion for protostellar masses $\gtrsim
10\:M_\odot$. Protostellar outflows, such as those included by {\it
  Wang et al.} (2010), also impede local accretion
to a star from some directions around the accretion radius.  This
issue is examined further in \S\ref{S:feedback}.

In sum, the key distinction between Competitive and Core Accretion is
whether competitive, ``clump-fed'' accretion of gas onto stars,
especially intermediate and massive stars, dominates over that present
in the initial pre-stellar core (PSC). In Core Accretion, the PSC
will likely gain some mass via accretion from the clump, but it will
also lose mass due to feedback; the net result is that the mass of the
PSC will be $\ga m_{*f}$. In Competitive Accretion, the PSC mass is
$\ll m_{*f}$.  Of course, reality may be somewhere between these
extremes, or might involve different aspects.  We note that an
observational test of this theoretical distinction requires that it be
possible to identify PSCs that may themselves be turbulent. As
discussed in \S\ref{S:numerical}, to date no simulations have been
performed with self-consistent initial conditions and with the full
range of feedback. Such simulations will be possible in the near
future and should determine whether massive PSCs can form in such an
environment, as required for Core Accretion models, or whether
low/intermediate mass stars can accrete enough mass by tidally
truncated Bondi accretion to grow into massive stars.

\subsubsection{Protostellar Collisions}

{\it Bonnell et al.} (1998) proposed massive stars may form
(i.e., gain significant mass) via protostellar collisions, including
those resulting from the hardening of binaries ({\it Bonnell and
  Bate,} 2005). 
This model was motivated by the perceived difficulty of accreting
dusty gas onto massive protostars---merging protostars are optically
thick and so unaffected by radiation pressure feedback. Note, such
protostellar collisions are distinct from those inferred to be driven
by binary stellar evolution ({\it Sana et al.,} 2012).  Universal
formation of massive stars via collisions would imply massive stars
always form in clusters.  Indeed, for collisional growth to be rapid
compared to stellar evolution timescales requires cluster
environments of extreme stellar densities, $\gtrsim 10^8\:{\rm
  pc^{-3}}$ (i.e., $n_{\rm H} \gtrsim 3\times 10^{9}\:{\rm cm^{-3}}$)
(e.g., {\it Moeckel and Clarke,} 2011; {\it Baumgardt and Klessen},
2011), never yet observed (Fig. 1).  {\it Moeckel and
  Clarke} (2011) find that when collisions are efficient,
they lead to runaway growth of one or two extreme objects, rather than
smoothly filling the upper IMF. Thus collisional growth
appears to be unimportant in typical massive star-forming
environments.

\subsection{Accretion Disks and Protostellar Evolution}\label{S:protostar}

\begin{figure}[bt]
\includegraphics[width=8cm]{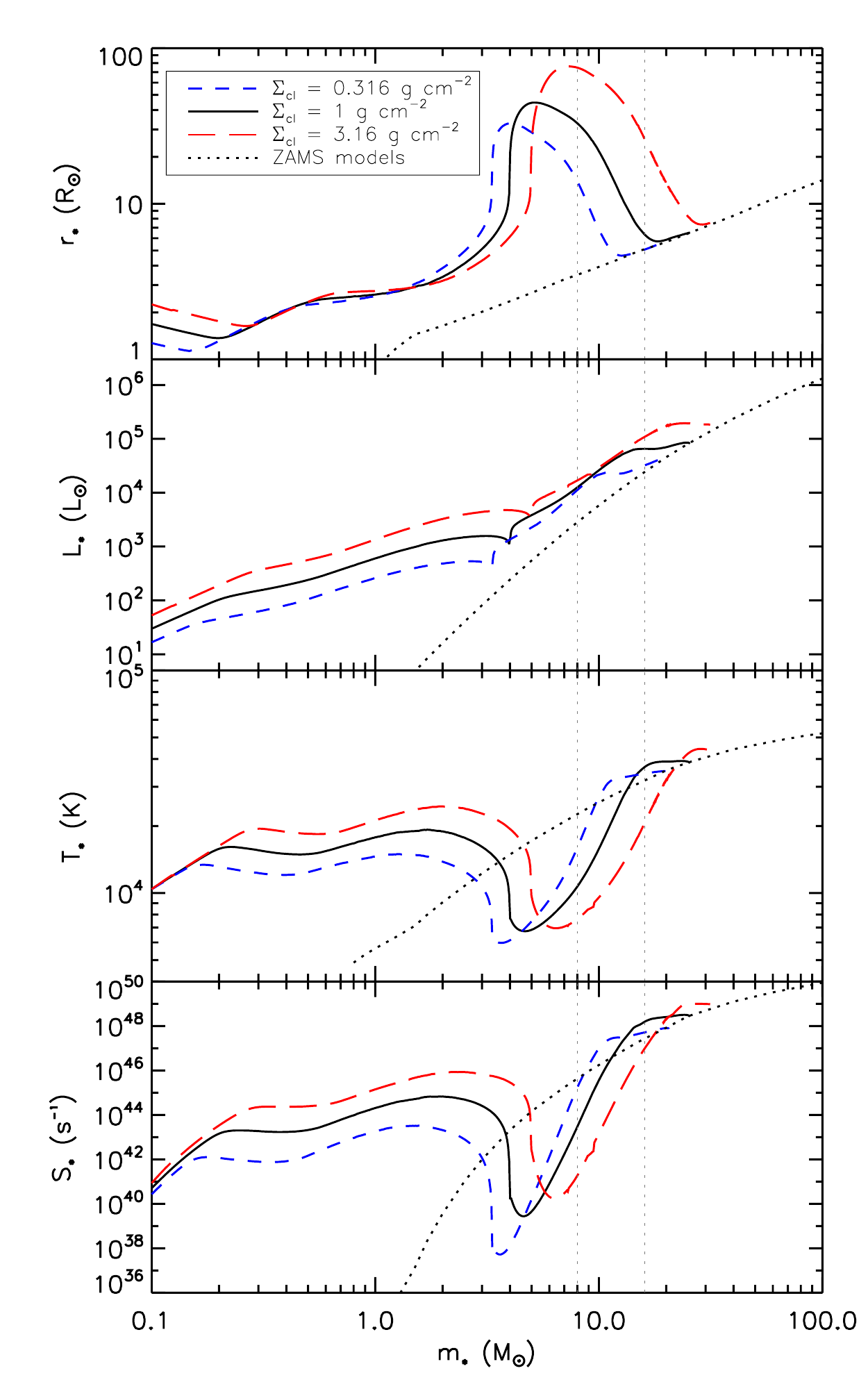}
 \caption{\small \label{fig:evolution} 
Evolution of a massive protostar forming from a $60\:M_\odot$ core in
$\Sigma_{\rm cl}\simeq 0.3, 1, 3\:{\rm g\:cm^{-2}}$ clumps.  Top to
bottom: radius, luminosity (including accretion),
surface temperature and H-ionizing luminosity ({\it Zhang et
  al.,} 2014; see also {\it Hosokawa et al.,} 2010). Dotted
lines show the zero age main sequence (ZAMS).
}
\end{figure}

\begin{figure*}[bt]
\begin{center}
\includegraphics[width=6.8in,angle=0]{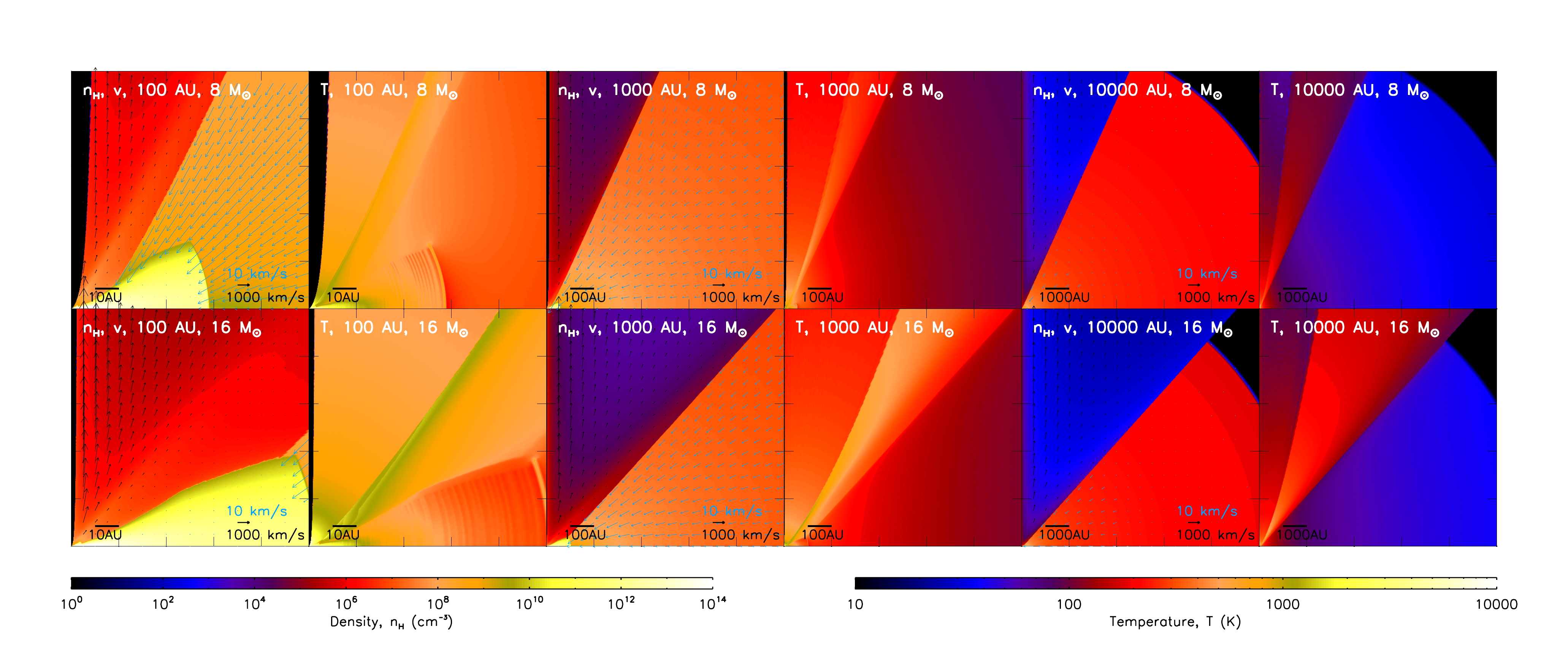}
 \caption{
\small \label{fig:yichen} 
Density and temperature profiles for a massive protostellar core ({\it
  Zhang et al.,} 2014), when the central protostar (at bottom left of
each panel) reaches $m_*=8\:M_\odot$ (top row) and 16~$M_\odot$
(bottom row). The disk midplane coincides with the x-axis; the
outflow/rotation axis with the y-axis. The core has initial mass
$M_c=60\:M_\odot$ and rotational to gravitational energy ratio of
$\beta_{\rm rot}=0.02$ and is embedded in a clump with mean surface density
$\Sigma_{\rm cl}=1\:{\rm g\:cm^{-2}}$. At each stage, three pairs of
box sizes are shown (left to right, 100, 10$^3$, 10$^4$~AU). Overlaid
on density plots are blue/black arrows showing infall/outflow
velocities (arrow length scale is 10/1000~$\rm km\:s^{-1}$,
respectively).}
\end{center}
\end{figure*}

In both Core and Competitive Accretion, the angular momentum of the
gas is expected to be large enough that most accretion to the
protostar proceeds via a disk. Here angular momentum is transferred
outwards via viscous torques resulting from the magneto-rotational
instability (MRI) or gravitational instability, which produces spiral
arms and, if strong enough, fragmentation to form a binary or higher
order multiple ({\it Kratter et al.,} 2010). For the Turbulent Core
model, an upper limit for the size of the disk forming in a core of
rotational energy $\beta_{\rm rot} =E_\mathrm{rot}/|E_\mathrm{grav}|$
is evaluated by assuming conservation of angular momentum of gas
streamlines inside the sonic point of the flow. Then the disk radius,
$r_d$, is a fraction of the initial core size: for a 60~$M_\odot$,
$\beta_{\rm rot}=0.02$ core forming in a clump with $\Sigma_{\rm cl}=1\:{\rm
  g\:cm^{-2}}$, we have $r_d=57.4, 102$~AU when $m_*=8, 16\:M_\odot$ ({\it
  Zhang et al.,} 2014; see Figs.~\ref{fig:evolution} \&
\ref{fig:yichen}). However, magnetic braking of the accretion flow may
make the disk much smaller ({\it Z.-Y. Li et al.,} this
volume). Disks around massive protostars also arise in
Competitive Accretion models (e.g., {\it Bate}, 2012), but are likely
to be smaller than in Core Accretion models.

Angular momentum may also be transferred via torques associated with a
large-scale magnetic field threading the disk that couples to a disk
wind ({\it Blandford and Payne,} 1982; {\it K\"onigl and Pudritz,}
2000). The final accretion to the protostar may be mediated by a
strong stellar $B$-field, as proposed for X-wind models around
low-mass protostars ({\it Shu et al.,} 2000). For massive stars, the
required field strengths would need to be $\gtrsim$kG ({\it Rosen et
  al.,} 2012). Or, the disk may continue all the way in to the
protostellar surface, in which case one expects high (near break-up)
initial rotation rates of massive stars. However, such high rates are
typically not observed and the necessary spin down would require
either stronger $B$-fields or longer disk lifetimes than those
inferred from observations ({\it Rosen et al.,} 2012).

The evolution of the protostar depends on its rate of accretion of
mass, energy and angular momentum from the disk. Since the dynamical
time of the star is short compared to the growth time, this process is
typically modeled as a sequence of equilibrium stellar structure
calculations (e.g., {\it Palla and Stahler,} 1991; {\it Hosokawa et
  al.,} 2010; {\it Kuiper and Yorke,} 2013; {\it Zhang et al.,}
2014). Most models developed so far have been for non-rotating
protostars (see {\it Haemmerl\'e et al.,} 2013 for an exception). A
choice must also be made for the protostellar surface boundary
condition: photospheric or non-photospheric. In the former, accreting
material is able to radiate away its high internal energy that has
just been produced in the accretion shock, while in the latter the gas
is optically thick (i.e., the photosphere is at a larger radius than
the protostellar surface). At a given mass, the protostar will respond
to advecting more energy by having a larger size. If accretion
proceeds through a disk, this is usually taken to imply photospheric
boundary conditions (cf., {\it Tan and McKee,} 2004). In the calculation shown in
Fig.~\ref{fig:evolution}, transition from quasi-spherical,
non-photospheric accretion is made to photospheric at $m_*\lesssim
0.1\:M_\odot$ based on an estimate of when outflows first affect the
local environment. Subsequent evolution is influenced by D-burning,
``luminosity-wave'' swelling and contraction to the zero age main
sequence (ZAMS) once the protostar is older than its current Kelvin-Helmholtz
time. This explains why protostars with higher accretion rates, i.e.,
in higher $\Sigma_{\rm cl}$ clumps, reach the ZAMS at higher
masses. Protostars may still accrete along the ZAMS. The high
temperatures and ionizing luminosities of this phase are a qualitative
difference from lower-mass protostars, especially leading to
ionization of the outflow cavity and eventually the core envelope
(\S\ref{S:feedback}).

Radiative transfer (RT) calculations are needed to predict the
multiwavelength appearance and total spectral energy distribution
(SED) of the protostar (e.g., {\it Robitaille et al.,} 2006; {\it
  Molinari et al.,} 2008; {\it Johnston et al.,} 2011; {\it Zhang and
  Tan,} 2011; {\it Zhang et al.,} 2013a; 2014). The luminosity of the
protostar and disk are reprocessed by the surrounding gas and dust in
the disk, envelope and outflow cavity. Figure~\ref{fig:yichen} shows
example models of the density, velocity and temperature structure of a
massive protostar forming inside a 60~$M_\odot$ core embedded in a
$\Sigma_{\rm cl}=1\:{\rm g\:cm^{-2}}$ environment at two stages, when
$m_*=8$ and $16\:M_\odot$, zooming from the inner 100 to 10$^3$ to
10$^4$~AU ({\it Zhang et al.,} 2014). One feature of these models is
that they self-consistently include the evolution of the protostar
from the initial starless core in a given clump environment, including
rotating infall envelope, accretion disk and disk-wind-driven outflow
cavities, that gradually open up as the wind momentum flux becomes
more powerful (\S\ref{S:feedback}). These models are still highly
idealized, being axisymmetric about a single protostar. A real source
would most likely be embedded in a forming cluster that includes other
protostars in the vicinity.

Similar continuum RT calculations have yet to be made for Competitive
Accretion. We anticipate they would show more disordered morphologies
and have smaller masses and $\Sigma$s of gas in the close vicinity of
the protostar than Core Accretion models, which may affect their SEDs
at a given evolutionary stage, i.e., value of $m_*$. These potential
SED differences are worth exploring, especially at the stages when
ionization becomes important for creating hypercompact (HC) and
ultracompact (UC) \ion{H}{2} regions (\S\ref{S:feedback}).

\vspace{-0.01in}
\subsection{Feedback Processes During Accretion}\label{S:feedback}

Massive protostars are much more luminous and hotter than low-mass
protostars so, all else being equal, one expects radiative feedback
(i.e., thermal heating, dissociation/ionization of hydrogen, radiation
pressure on dust) to be more important. The same is true for
mechanical feedback from stellar winds (i.e., those from the stellar
surface) and protostellar outflows (magneto-centrifugally-driven flows
powered by accretion). Alternatively, if massive stars tend to form in
denser, more highly-pressurized environments, then feedback will have
a harder time disrupting accretion. For Core Accretion models, a major
goal of feedback studies is to estimate the star formation efficiency
from a core of a given initial mass and density (or surrounding clump
pressure). For both Core and Competitive Accretion, an important goal
is to determine whether there are processes that lead to IMF
truncation at some maximum mass. Feedback may also affect the ability
of a core to fragment to form a binary and the efficiency of a clump
to fragment into a cluster. Feedback also produces observational
signatures, such as outflow cavities, \ion{H}{2} regions and
excitation of masers, that all serve as diagnostics of massive star
formation. A general review of feedback is given by {\it Krumholz et
  al.} in this volume. Here we discuss processes directly relevant to
massive star formation.

For massive stars to form from massive cores, a mechanism is needed to
prevent the core from fragmenting. {\it Krumholz and Mckee} (2008)
suggested this may be due to radiative feedback from surrounding
lower-mass protostars that have high accretion luminosities if they
are forming in a high pressure clump. This model predicts a minimum
$\Sigma$ for clumps to form massive stars, $\Sigma_{\rm cl}\gtrsim
1\:{\rm g\:cm^{-2}}$. On the other hand, {\it Kunz and Mouschovias}
(2009) and {\it Tan et al.} (2013b) invoked a non-feedback-related
mechanism of magnetic field support to allow massive cores to resist
fragmentation. This does not require a minimum $\Sigma_{\rm cl}$
threshold, but does require that there be relatively strong, $\sim$mG,
$B$-fields in at least some parts of the clump, so that the core mass
is set by the magnetic critical mass. Simulations confirm that
magnetic fields can suppress fragmentation (\S\ref{S:numerical}). The
observational evidence for a whether there is a $\Sigma_{\rm cl}$
threshold for massive star formation is discussed in
\S\ref{S:conditions}.

Once a massive protostar starts forming, but before contraction to the
ZAMS, the dominant feedback is expected to be due to protostellar
outflows (see also \S\ref{S:outflows}). As a consequence of their
extraction of angular momentum, these magneto-centrifugally-launched
disk- and/or X-winds tend to have mass flow rates
$\dot{m}_w=f_w\dot{m}_*$ with $f_w\sim0.1-0.3$ and terminal velocities
$v_w\sim v_K(r_0)$, where $v_K$ is the Keplerian speed in the disk at
the radius, $r_0$, of the launching region. The total outflow momentum
flux can be expressed as $\dot{p}_w = \dot{m}_w v_w = f_w\dot{m}_* v_w
\equiv \phi_w\dot{m}_* v_K(r_*)$: {\it Najita and Shu} (1994) X-wind
models have dimensionless parameter $\phi_w \simeq 0.6$.  An
implementation of the {\it Blandford and Payne} (1982) disk-wind model
has $\phi_w \simeq 0.2$, relatively independent of $m_*$ ({\it Zhang
  et al.,} 2013a). Outflows are predicted to be collimated with
$dp_w/d\Omega = (p_w/4\pi)[{\rm
    ln}(2/\theta_0)(1+\theta_0^2-\mu^2)]^{-1}$ ({\it Matzner and
  McKee,} 1999), where $\mu={\rm cos} \theta$, $\theta$ is the angle
from outflow axis, and $\theta_0\sim 10^{-2}$ is a small angle of the
core of the outflow jet. {\it Matzner and McKee} (2000) found star
formation efficiency from a core due to such outflow momentum feedback
of $\bar{\epsilon}_{*d}\sim 0.5$. For the protostars in
Fig.~\ref{fig:evolution}, $\bar{\epsilon}_{*d}\simeq 0.45, 0.57, 0.69$
for $\Sigma_{\rm cl}=0.3, 1, 3\:{\rm g\:cm^{-2}}$, indicating
protostellar outflow feedback may set a relatively constant formation
efficiency from low to high mass cores.

The protostar's luminosity heats its surroundings, mostly via
absorption by dust, which at high densities ($n_{\rm H}\gtrsim
10^5\:{\rm cm^{-3}}$) is well-coupled thermally to the gas ({\it Urban
  et al.,} 2010). Dust is destroyed at $T\gtrsim 1500$~K, i.e., at
$\lesssim 10$~AU for models in Fig.~\ref{fig:yichen}. Hot core
chemistry (\S\ref{S:hotcore}) is initiated for temperatures $\gtrsim
100$~K. Thermal heating reduces subsequent fragmentation in the disk
(see \S\ref{S:numerical}).

As the protostar grows in mass and settles towards the main sequence,
the temperature and H-ionizing luminosity begin to increase. The
models in Fig.~\ref{fig:yichen} have H-ionizing photon luminosities of
$2.9\times 10^{43}\:{\rm s}^{-1}$ and $1.6\times 10^{48}\:{\rm
  s^{-1}}$ when $m_*= 8\:M_\odot$ and $16\:M_\odot$, respectively. A
portion of the inner outflow cavity will begin to be ionized---an
``outflow-confined'' \ion{H}{2} region ({\it Tan and McKee,} 2003).
Inner, strongly-bound parts of the infall envelope that are unaffected
by outflows could also confine the \ion{H}{2} region ({\it Keto,}
2007). The \ion{H}{2} region structure is detectable via radio
continuum observations of thermal bremsstrahlung emission
(\S\ref{S:HII}). Its extent depends sensitively on the density and
dust content of the gas. Feedback from the \ion{H}{2} region is driven
by its high temperature, $\sim 10^4$~K, that sets up a pressure
imbalance at the ionization front boundary with neutral gas. Since the
MHD-outflow momentum flux is likely to dominate over the \ion{H}{2}
region thermal pressure, ionization feedback will only begin to be
effective once the entire outflow cavity is ionized and ionization
fronts start to erode the core infall envelope (cf., {\it Peters et
  al.,} 2011). Once the core envelope is mostly cleared, leaving
equatorial accretion and a remnant accretion disk, the diffuse
ionizing radiation field that is processed by the disk atmosphere can
photoevaporate the disk ({\it Hollenbach et al.,} 1994). This process
has been invoked by {\it McKee and Tan} (2008) to shut off accretion
of the first stars around $\sim 100-200\:M_\odot$ (see also {\it
  Hosokawa et al.,} 2011; {\it Tanaka et al.}  2013), but its role in
present-day massive star formation is unclear, especially given the
presence of dust that can absorb ionizing photons.

Radiation pressure acting on dust has long been regarded as a
potential impediment to massive star formation ({\it Larson \&
  Starrfield}, 1971; {\it Kahn,} 1974; {\it Wolfire and Cassinelli,}
1987). However, as long as the accretion flow remains optically thick,
e.g., in a disk, then there does not seem to be any barrier to forming
massive stars ({\it Nakano,} 1989; {\it Jijina and Adams,} 1996; {\it
  Yorke and Sonnhalter,} 2002; {\it Krumholz et al.,} 2009; {\it
  Kuiper et al.,} 2010a; 2011; {\it Tanaka and Nakamoto,}
2011). Outflows also reduce the ability of radiation pressure to
terminate accretion, since they provide optically thin channels
through which the radiation can escape ({\it Krumholz et al.,}
2005b). This contributes to the ``flashlight effect'' ({\it Yorke and
  Sonnhalter,} 2002; {\it Zhang et al.,} 2013a), leading to factors of
several variation in the bolometric flux of a protostar depending on
viewing angle.  Numerical simulations of radiation pressure feedback
are summarized in \S\ref{S:numerical}.

A potential major difference between Core and Competitive Accretion is
their ability to operate in the presence of feedback. As discussed
above, core accretion to a disk is quite effective at resisting
feedback: gas comes together into a self-gravitating object before the
onset of star formation.  Competitive accretion of ambient gas from
the clump may be more likely to be disrupted by feedback. Considering
the main feedback mechanism for low-mass stars, we estimate the
ram pressure 
associated with 
a MHD (X- or disk-) wind of mass-loss
rate $\dot{m}_w$ and velocity $v_w$ as $P_{w} = \rho_w v_w^2 =
f_\theta \dot{m}_w v_w/(4\pi r^2) = f_\theta \phi_w \dot{m}_*
v_K(r_*)/(4\pi r^2)$, where $f_\theta \equiv 0.1 f_{\theta,0.1}$ is
the factor by which the momentum flux of the wide-angle component of
the wind is reduced from the isotropic average and where we have
normalized, conservatively, to parameter values implied by disk-wind
or X-wind models (e.g., the fiducial {\it Matzner and McKee,} 1999
distribution has a minimum $f_\theta\simeq 0.2$ at
$\theta=90^\circ$). Evaluating $P_{w}$ at $r_{\rm BH} =
2Gm_*/\sigma^2$ (appropriate for competitive accretion from a
turbulent clump) around a protostar of current mass $m_*\equiv
m_{*,1}M_\odot$, adopting accretion rates from
eq.(\ref{eq:mdot_comp2}) and setting $r_* \equiv r_{*,3} 3 R_\odot$
(Fig.~\ref{fig:evolution}), we find the condition for the clump mean
pressure to overcome the ram pressure of the wind, $\bar{P}_{\rm
  cl}>P_{w}(r_{\rm BH})$:
\beq
\Sigma_{\rm cl}> 11.7 \left( \frac{ f_{\theta,0.1} \phi_{w,0.1}\alpha_{\rm vir} \epsilon_{\rm ff,0.1}}{\phi_B\phi_{\rm geom}\epsilon_{\rm cl,0.5}} \right)^4 \frac{M_{\rm cl,3}^3 m_{*f,1}^4}{r_{*,3}^{2}m_{*,1}^{6}}\:{\rm g\:cm^{-2}}\label{eq:comp_condition}
\eeq
Thus in most clumps shown in Fig.~\ref{fig:overview}, $\bar{P}_{\rm
  cl}$ 
is too weak to confine gas inside the Bondi radius in the
presence of such outflows. Note, here $m_*$ is the mass scale at which
feedback is being considered, while $m_{*f}$ parameterizes the
accretion rate needed to form a star of final mass
$m_{*f}$. Eq.~(\ref{eq:comp_condition}) shows MHD-wind feedback
generated by the accretion rates expected in
Competitive Accretion 
severely impacts accretion over most of the
Bondi-sphere, especially if the mass scale at which competitive
accretion starts, following initial core accretion, is small ($m_*\sim
1\:M_\odot$).  We suggest simulations have so far not fully 
resolved the effects of MHD-wind feedback on competitive accretion
and that this feedback may lead to a major reduction in its efficiency.

\subsection{Results from Numerical Simulations}\label{S:numerical}

\begin{figure*}
\epsscale{1.8}
\plotone{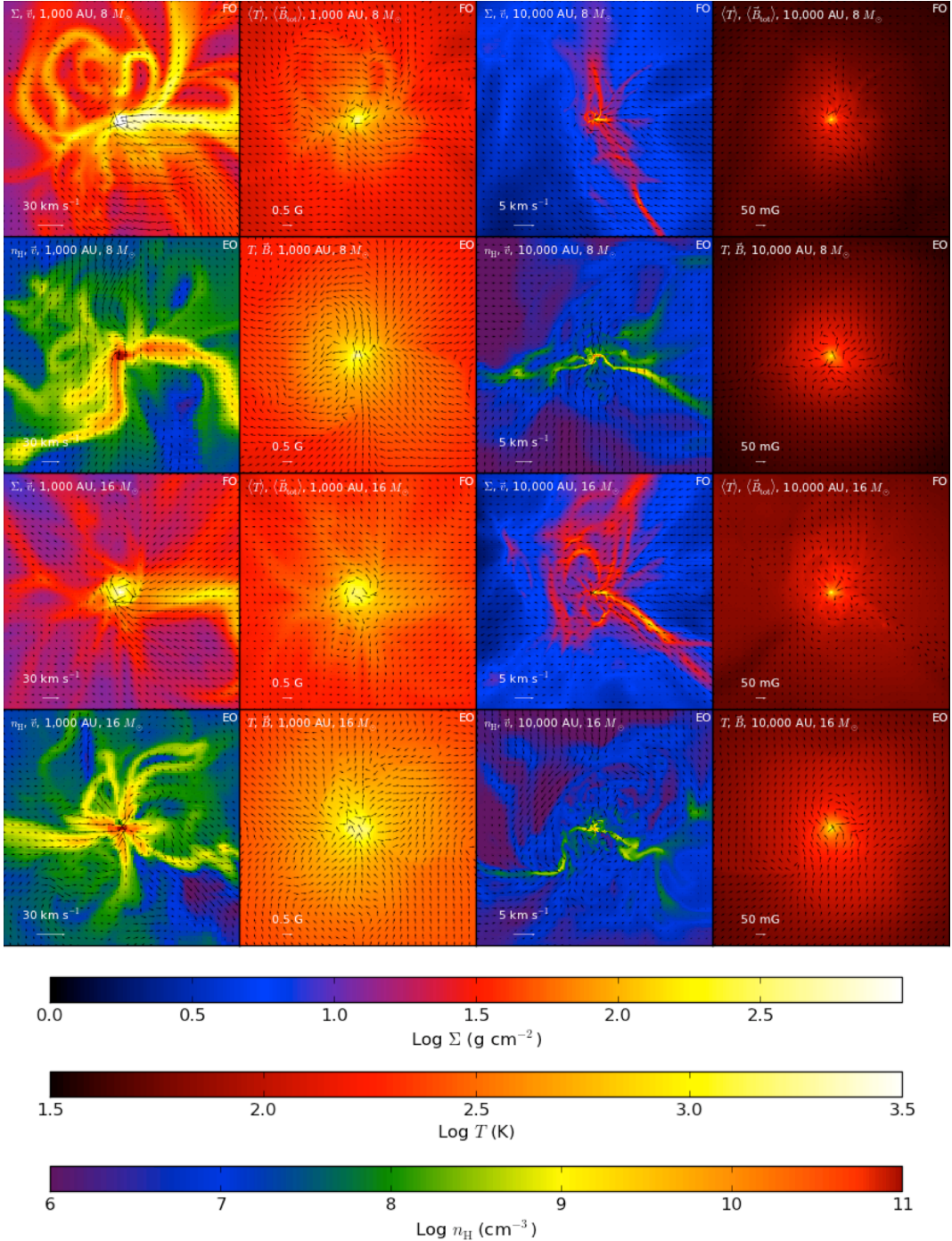}
\caption{\small \label{fig:myers} 
Simulation of massive star formation including MHD and radiation
pressure feedback ({\it Myers et al.,} 2013) from an initial core with
$M_c=300\:M_\odot\simeq 2 M_\Phi$ (i.e., twice the magnetic critical
mass), $\Sigma_c=2\:{\rm g\:cm^{-2}}$ and $\sigma=2.3\:{\rm
  km\:s^{-1}}$ (so that $\alpha_{\rm vir}=2.1$; turbulence decays,
leading to global collapse). Protostellar outflow feedback (e.g., {\it
  Wang et al.,} 2010; {\it Cunningham et al.}, 2011) is not included.
The top row shows a face-on (FO) view of the accretion disk centered
on the protostar when it has a mass of $8\:M_\odot$: from left to
right are: mass surface density, $\Sigma$, and mean velocity, $\langle
v \rangle$, (arrows) in a 10$^3$~AU box; mass-weighted temperature,
$\langle T \rangle$, and total magnetic field strength, $\langle
B_{\rm tot}\rangle$, (arrows) in a 10$^3$~AU box; then the same two
figures but for a 10$^4$~AU box. The second row shows edge-on (EO)
views of this structure, with slices in a plane containing the
protostar of, from left to right: H number density, $n_{\rm H}$ and
velocity, $v$, (arrows) of a 10$^3$~AU square; temperature, $T$, and
in-plane component of magnetic field, $B$, (arrows) of a 10$^3$~AU
square; then the same two figures but for a 10$^4$~AU square. The
third row repeats the first row, but now for a $16\:M_\odot$
protostar, and the fourth row repeats the second row for this
protostar.  With an initially turbulent core, the accretion flow is
relatively disordered (cf., Fig.~\ref{fig:yichen}).
}
\end{figure*}

Numerical simulations have long been a major tool for investigating
massive star formation. Today, the majority use either the Lagrangian
technique smoothed particle hydrodynamics (SPH; e.g., {\it Lucy,}
1977) or the Eulerian technique adaptive mesh refinement (AMR; {\it
  Berger and Oliger,} 1984), both of which provide high dynamic range
allowing collapse to be followed over orders of magnitude in length
scale in general geometries.  Both code types include self-gravity,
hydrodynamics and sink particles to represent stars
(e.g., {\it Bate et al.,} 1995; {\it Krumholz et al.,} 2004).

Probably the most significant advance in simulations since Protostars
\& Planets V has been addition of extra physical processes.  For SPH,
there are implementations of magnetohydrodynamics ({\it Price and
  Monaghan,} 2004), flux-limited diffusion (FLD) for RT of
dust-reprocessed non-ionizing radiation ({\it Whitehouse and Bate,}
2004), and ray-tracing for ionizing RT ({\it Dale et al.,} 2005; {\it
  Bisbas et al.,} 2009), the latter specifically used to study massive
star formation.  AMR codes include an even broader range of physics,
all of which has been brought to bear on massive star formation: MHD
(e.g., {\it Fryxell et al.,} 2000; {\it Li et al.,} 2012a), FLD for
non-ionizing radiation ({\it Krumholz et al.,} 2007b; {\it Commer{\c
    c}on et al.,} 2011b), ray-tracing for ionizing and (in restricted
circumstances) non-ionizing radiation ({\it Peters et al.,} 2010a),
and protostellar outflows ({\it Wang et al.,} 2010; {\it Cunningham et
  al.,} 2011). More sophisticated RT schemes than pure FLD or
ray-tracing are also available in some non-adaptive grid codes (e.g.,
{\it Kuiper et al.,} 2010b).

While this is progress, a few caveats are in order. First, no code yet
includes all of these physical processes: e.g., ORION includes MHD,
dust-reprocessed radiation and outflows, but not ionizing radiation,
while FLASH has MHD and ionizing radiation, but not outflows or
dust-reprocessed radiation. Second, some physical processes have only
been studied in isolation by a single code and their relative
importance is unclear. Examples include imperfect thermal coupling
between gas and dust ({\it Urban et al.,} 2010), dust coagulation and
drift relative to gas ({\it Suttner et al.,} 1999), and ambipolar
drift ({\it Duffin and Pudritz,} 2008).

Still, the advances in simulation technique have yielded some
important general conclusions. First, concerning fragmentation,
hydrodynamics-only simulations found that collapsing gas clouds
invariably fragmented into stars with initial masses of $\sim 0.1$
$M_\odot$ ({\it Bonnell et al.,} 2004; {\it Dobbs et al.,} 2005),
implying formation of massive stars would have to arise via subsequent
accretion onto these fragments. However, {\it Krumholz et al.} (2007a,
2012) showed that adding non-ionizing, dust reprocessed radiative
feedback suppresses this behavior, as the first few stars to form heat
the gas around them via their accretion luminosities, raising the
Jeans mass and preventing much fragmentation.  Similarly, {\it
  Hennebelle et al.} (2011) showed that magnetic fields also inhibit
fragmentation, and {\it Commer{\c c}on et al.} (2011a) and {\it Myers
  et al.} (2013; Fig.~\ref{fig:myers}) have combined magnetic fields
and radiation to show that the two together suppress fragmentation
much more effectively than either one alone.

Second, massive star feedback is not very effective at halting
accretion. Photoionization can remove lower density gas, but dense gas
that is already collapsing onto a massive protostar is largely
self-shielding and is not expelled by ionizing radiation ({\it Dale et
  al.,} 2005; {\it Peters et al.,} 2010a; b; 2011). As for radiation
pressure, 2D simulations with limited resolution ($\sim 100$ AU)
generally found that it could reverse accretion, thus limiting final
stellar masses ({\it Yorke and Sonnhalter,} 2002). However, higher
resolution 2D and 3D simulations find that radiation pressure does not
halt accretion since, in optically thick flows, the gas is capable of
reshaping the radiation field and beaming it away from the bulk of the
incoming matter.  This beaming can be provided by radiation
Rayleigh-Taylor fingers (e.g., {\it Krumholz et al.,} 2009; {\it Jiang
  et al.,} 2013; cf., {\it Kuiper et al.,} 2012), by an optically
thick disk ({\it Kuiper et al.,} 2010a, 2011; {\it Kuiper and Yorke,}
2013), or by an outflow cavity ({\it Krumholz et al.,} 2005b; {\it
  Cunningham et al.,} 2011).

While there is general agreement on the two points above, simulations
have yet to settle the question of whether stars form primarily via
Competitive or Core Accretion, or some hybrid of the two. Resolving
this requires simulations large enough to form an entire star cluster,
rather than focusing on a single massive star. {\it Bonnell et al.}
(2004) and {\it Smith et al.} (2009)
tracked the mass that eventually ends up in massive stars in their
simulations of cluster formation, concluding that it is drawn from a
$\sim 1$~pc cluster-sized region rather than a single well-defined
$\sim 0.1$~pc core, and that there are no massive bound structures
present. However, these simulations lacked radiative feedback or
$B$-fields, and thus likely suffer from over-fragmentation.
{\it Wang et al.} (2010) performed simulations including outflows and
magnetic fields. They found the most massive star in their
simulations ultimately draws its mass from a $\sim 1$ pc-sized
reservoir comparable to the size of its parent cluster, consistent
with Competitive Accretion, but that the flow on large scales is
mediated by outflows, preventing onset of rapid global collapse.
As described in \S\ref{S:competitive}, the average accretion rate to
the massive star is relatively low in this simulation compared to the
expectations of Core Accretion, which may be possible to test
observationally (\S\ref{S:accretion}). As discussed in
\S\ref{S:feedback}, outflow feedback on the scale of the Bondi
accretion radius may be important in further limiting the rate of
competitive accretion and so far has not been well resolved.

{\it Peters et al.} (2010a; b) simulated massive cluster formation
with direct (i.e., not dust-reprocessed) radiation and $B$-fields,
starting from smooth, spherical, slowly-rotating initial
conditions. They found massive stars draw mass from large but
gravitationally-bound regions, but that the mass flow onto these stars
is ultimately limited by fragmentation of the accreting gas into
smaller stars. {\it Girichidis et al.} (2012) extended this result to
more general geometries. {\it Krumholz et al.} (2012) conducted
simulations of cluster formation including radiation and starting from
an initial condition of fully-developed turbulence. They found massive
stars do form in identifiable massive cores, with several tens of
solar masses within $\sim 0.01$~pc. Core mass is not conserved in a
Lagrangian sense, as gas flows in or out, but they are nonetheless
definable objects in an Eulerian sense.

These contradictory results likely have several origins. One is
initial conditions (e.g., {\it Girichidis et al.,} 2012). Those
lacking any density structure and promptly undergoing global collapse
(e.g., {\it Bonnell et al.,} 2004) tend to find there are no bound,
massive structures that can be identified as the progenitors of
massive stars, while those either beginning from saturated turbulence
(e.g., {\it Krumholz et al.,} 2012) or self-consistently producing it
via feedback (e.g., {\it Wang et al.,} 2010) do contain structures
identifiable as massive cores. Another issue is the different range of
included physical mechanisms, with none of the published cluster
simulations combining dust-reprocessed radiation and magnetic
fields---shown to be so effective at suppressing fragmentation on
smaller scales. A final issue may simply be one of interpretation,
with SPH codes tending to focus on the Lagrangian question of where
the individual mass elements that make up a massive star originate,
while Eulerian codes focus on the presence of structures at a
particular point in space regardless of the paths of individual fluid
elements.

\section{\textbf{OBSERVATIONS OF INITITAL CONDITIONS}}\label{S:initial}

\vspace{-0.01in}
\subsection{Physical Properties of Starless Cores \& Clumps}\label{S:physical_starless}

Initial conditions in Core Accretion models are massive starless
cores, with $\Sigma \sim 1\:\gcc$, similar to the $\Sigma$s of their
natal clump. For Competitive Accretion, massive stars are
expected to form near the centers of the densest clumps. Thus to test
these scenarios, methods are needed to study high $\Sigma$ and volume
density ($n_{\rm H}\sim 10^6\:{\rm cm^{-3}}$), compact ($r_c\sim
0.1$~pc, i.e., 7\arcsec\ at 3~kpc) and potentially very cold ($T\sim
10$~K) structures. Recently, many studies of initial conditions have
focussed on Infrared Dark Clouds (IRDCs): regions with such high
$\Sigma$ that they appear dark at MIR ($\sim 10\:{\rm \mu m}$) and
even up to FIR ($\sim 100\:{\rm \mu m}$) wavelengths. Indeed,
selection of cores that may be starless often involves checking for
the absence of a source at 24 or 70~$\rm \mu m$.

\subsubsection{Mass Surface Densities, Masses \& Temperatures}

One can probe $\Sigma$ structures via MIR extinction mapping of IRDCs,
using diffuse Galactic background emission from warm dust. {\it
  Spitzer} IRAC (e.g., GLIMPSE; {\it Churchwell et al.,} 2009) $8\:{\rm
  \mu m}$ images resolve down to $2''$ and can probe to $\Sigma \sim
0.5\:{\rm g\:cm^{-2}}$ (e.g., {\it Butler and Tan,} 2009; 2012 [BT12];
{\it Peretto and Fuller,} 2009; {\it Ragan et al.,} 2009). The method
depends on the $8\:{\rm \mu m}$ opacity per unit total mass,
$\kappa_{\rm 8\mu m}$ ({\it BT12} use $7.5\:{\rm cm^{2}\:g^{-1}}$ based on
the moderately coagulated thin ice mantle dust model of {\it Ossenkopf
  and Henning,} 1994), but is independent of dust temperature,
$T_d$. Allowance is needed for foreground emission, best measured by
finding ``saturated'' intensities towards independent, optically thick
cores ({\it BT12}). Only differences in $\Sigma$ relative to local
surroundings are probed, so the method is insensitive to low-$\Sigma$
IRDC environs. This limitation is addressed by combining NIR \& MIR
extinction maps ({\it Kainulainen and Tan,} 2013). Even with careful
foreground treatment, there are $\sim 30\%$ uncertainties in $\kappa$
and thus $\Sigma$, and, adopting 20\% kinematic distance
uncertainties, a 50\% uncertainty in mass. Ten IRDCs studied in this
way are shown in Fig.~\ref{fig:overview}.

The high resolution $\Sigma$ maps derived from {\it Spitzer} images
allow measurement of core and clump structure. Parameterizing density
structure as $\rho \propto r^{-k_\rho}$ and looking at 42 peaks in
their $\Sigma$ maps, {\it BT12} found $k_\rho\simeq 1.1$ for
``clumps'' (based on total $\Sigma$ profile) and $k_\rho\simeq 1.6$
for ``cores'' (based on $\Sigma$ profile after clump envelope
subtraction). These objects, showing total $\Sigma$, are also plotted
in Fig.~\ref{fig:overview}. A $\Sigma$ map of one of these core/clumps
is shown in Fig.~\ref{fig:shuo}. {\it Tan et al.}  (2013b) used the
fact that some of these cores are opaque at $70\:{\rm \mu m}$ to
constrain $T_d\lesssim 13$~K. {\it Ragan et al.}  (2009) measured an
IRDC core/clump mass function, ${\rm d} N/{\rm d} M \propto
M^{-\alpha_{\rm cl}}$ with $\alpha_{\rm cl} \simeq 1.76\pm 0.05$ from
30 to 3000~$M_\odot$, somewhat shallower than that of the Salpeter
stellar IMF ($\alpha_*\simeq 2.35$).

The $\Sigma$ of these clouds can also be probed by the intensity,
$S_\nu/\Omega$, of FIR/mm dust emission, requiring $T_d$
and $\kappa_\nu$. For optically thin RT and black body emission,
$\Sigma = 4.35 \times 10^{-3} ([S_\nu/\Omega] / [{\rm MJy/sr}])
\kappa_{\nu,0.01}^{-1} \lambda_{1.2}^3 [{\rm exp}(0.799 T_{d,15}^{-1}
  \lambda_{1.2}^{-1})-1]\:{\rm g\:cm^{-2}}$, where
$\kappa_{\nu,0.01}\equiv \kappa_{\nu}/[0.01{\rm cm^2/g}])^{-1}$,
$\lambda_{1.2}\equiv\lambda/1.2\:{\rm mm}$ and $T_{d,15}\equiv
T_d/15\:{\rm K}$. A common choice of $\kappa_\nu$ is again that
predicted by the moderately coagulated thin ice mantle dust model of
{\it Ossenkopf and Henning} (1994), with opacity per unit dust mass of
$\kappa_{{\rm 1.2mm},d}=1.056\:{\rm cm^2\:g^{-1}}$. A
gas-to-refractory-component-dust-mass ratio of 141 is estimated by
{\it Draine} (2011)
so $\kappa_{\rm 1.2mm}=7.44\times 10^{-3}\:{\rm cm^2\:g^{-1}}$.
Uncertainties in $\kappa_\nu$ and $T_d$ now contribute to $\Sigma$:
e.g., {\it Tan et al.} (2013b) adopt $\kappa$ uncertainties of 30\%
and $T_d=10\pm 3$~K, leading to factor $\sim 2$ uncertainties in
1.3~mm-derived $\Sigma$s. {\it Rathborne et al.}~(2006)
studied 1.2~mm
emission at 11\arcsec\ resolution in 38 IRDCs finding core/clumps
with $\sim 10$ to $10^4\:M_\odot$ (for $T_d=15$~K). In their sample of
140 sources they found ${\rm d} N/{\rm d} M
\propto M^{-\alpha_{\rm cl}}$, with $\alpha_{\rm cl}\simeq 2.1\pm 0.4$.

{\it Herschel} observations of dust emission at 70 to 500~$\rm \mu m$
allow simultaneous measurement of $T_d$ and $\Sigma$ at $\sim
20-30\arcsec$ resolution and numerous studies have been made of IRDCs
(e.g., {\it Peretto et al.,} 2010; {\it Henning et al.,} 2010; {\it
  Beuther et al.,} 2010a; {\it Battersby et al.,} 2011; {\it Ragan et
  al.,} 2012). For MIR-dark regions, {\it Battersby et al.} (2011)
derived a median $\Sigma\simeq 0.2\:{\rm g\:cm^{-2}}$, but with some
values extending to $\sim 5\:{\rm g\:cm^{-2}}$. The median $T_d$ of
regions with $\Sigma\gtrsim 0.4\:{\rm g\:cm^{-2}}$ was 19~K, but the
high-$\Sigma$ tail had $T_d\sim 10$~K.

Interferometric studies have probed mm dust emission at higher
resolution. ``Clumps'' are often seen to contain substructure, i.e., a
population of ``cores''.  CMF measurements have been attempted: e.g.,
{\it Beuther and Schilke} (2004; see also {\it Rod\'on et al.,} 2012)
observed IRAS 19410+2336, finding $\alpha_c = 2.5$ from $M_c \sim 1.7$
to $25\:M_\odot$ (but with few massive cores). While the similarity of
CMF and IMF shapes is intriguing, there are caveats, e.g., whether
cores are resolved; whether they are PSCs rather than non-star-forming
overdensities or already star-forming cores; the possibility of
mass-dependent lifetimes of PSCs ({\it Clark et al.,} 2007); and
binary/multiple star formation from PSCs.

Some massive ($\sim60\:M_\odot$) cores, e.g., IRDC G28.34+0.06 P1
({\it Zhang et al.,} 2009), Cygnus X N63 ({\it Bontemps et al.,} 2010;
note recent detection of a bipolar outflow indicates a protostar is
forming in this source, {\it Duarte-Cabral et al.,} 2013) and IRDC
C1-S ({\it Tan et al.,} 2013b) (Figs.~\ref{fig:overview} \&
\ref{fig:shuo}), have apparently monolithic, centrally-concentrated
structures with little substructure, even though containing many
($\sim 100$) Jeans masses. This suggests fragmentation is being
inhibited by a nonthermal mechanism, i.e., magnetic fields. {\it Tan
  et al.}  (2013b) estimate $\sim 1$~mG field strengths are needed for
the mass of C1-S to be set by its magnetic critical mass, given its
density of $\bar{n}_{\rm H}\simeq 6\times 10^5\:{\rm cm^{-3}}$.

Many molecular lines have been observed from IRDCs. Using integrated
molecular line intensities to derive $\Sigma$ is possible in theory,
but common species like CO are frozen-out from the gas phase (see
below), and other species still present have uncertain and likely
spatially varying abundances. Nevertheless, if the astrochemistry is
understood, then species that are expected to become relatively
abundant in the cold, dense conditions of starless cores, such as
deuterated N-bearing molecules (\S\ref{S:astrochem}), can be used to
identify PSCs, distinguishing them from the surrounding clump.

IRDC gas temperatures of 10--20 K have been derived from $\rm
NH_3$ inversion transitions (e.g., {\it Pillai et al.,} 2006; {\it
  Wang et al.,}~2008; {\it Sakai et al.,} 2008; {\it Chira et al.,}
2013).

\subsubsection{Magnetic Fields}

Polarization of dust continuum emission is thought to arise from
alignment of non-spherical grains with $B$-fields and is thus a
potential probe of plane-of-sky projected field morphology and, with
greater uncertainty, field strength. The correlated orientation of
polarization vectors with the orientations of filaments, together with
the correlated orientations of polarization vectors of dense cores
with their lower density surroundings ({\it H. Li et al.,} this
volume) suggests $B$-fields play some role in the formation of dense
cores. However, these polarization results are typically for
relatively nearby molecular clouds, such as Taurus, Pipe Nebula and
Orion, and only a few, lower-resolution studies have been reported for
IRDCs ({\it Matthews et al.,} 2009).

\begin{figure}[bt]
\includegraphics[width=8cm]{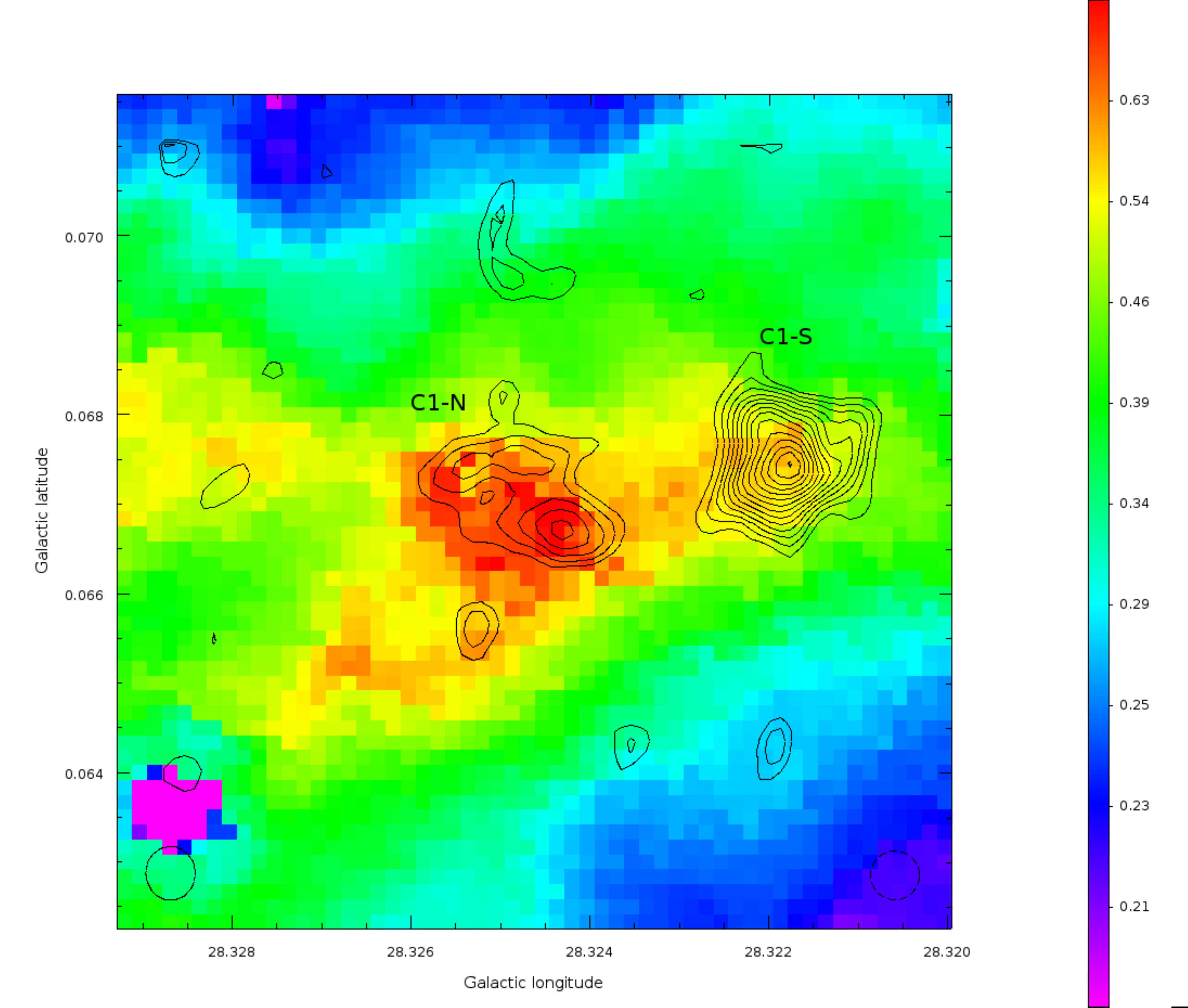}
 \caption{\small \label{fig:shuo}
Candidate massive starless cores, C1-S \& C1-N, traced by $\rm
N_2D^+$(3-2) (contours), observed by {\it ALMA} ({\it Tan et al.,}
2013b). Background shows MIR $\Sigma$ map ($\rm g\:cm^{-2}$).
C1-S has $\sim 60\:M_\odot$.
The high value of $\rm [N_2D^+]$/$\rm
[N_2H^+]\sim 0.1$ ({\it Kong et al.,} in prep.) is a chemical indicator that C1-S is starless.  }
\end{figure}

Line-of-sight $B$-field strengths can be derived from Zeeman splitting
of lines from molecules with an unpaired electron, such as OH, which
probes lower-density envelopes, and CN, which traces denser
gas. Unfortunately, measurement of Zeeman splitting in CN is very
challenging observationally, requiring bright lines, and the reported
measurements in massive star-forming regions are all towards already
star-forming cores (\S\ref{S:accretion}). From the results of {\it
  Falgarone et al.} (2008) as summarized by {\it Crutcher} (2012), at
densities $n_{\rm H}\gtrsim 300\:{\rm cm^{-3}}$, $B_{\rm max}\simeq
0.44 (n_{\rm H}/10^5\:{\rm cm^{-3}})^{0.65}$~mG, with a distribution
of $B$ flat between 0 and $B_{\rm max}$. Such field stengths, if
present in massive starless cores like C1-S (Fig.~\ref{fig:shuo}),
could support them against rapid free-fall collapse and fragmentation.

\subsubsection{Kinematics and Dynamics}

Measurement of cloud kinematics requires molecular line tracers, but
again one faces the problem of being sure which parts of the cloud
along the line of sight are being probed by a given tracer. The
kinematics of ionized and neutral species can differ due to magnetic
fields ({\it Houde et al.,} 2009).  The spectra of molecular tracers
of IRDCs, such as $^{13}$CO, C$^{18}$O, \NNH, \AMM, HCN, HCO$^+$, CCS,
show line widths $\sim 0.5 - 2$ \kms, i.e., consistent with varying
degrees of supersonic turbulence (e.g., {\it Wang et al.,}~2008; {\it
  Sakai et al.,}~2008; {\it Fontani et al.,}~2011).  In studying the
kinematics of IRDC G035.39–00.33, {\it Henshaw et al.} (2013) have
shown it breaks up into a few distinct filamentary components
separated by up to a few $\rm km\:s^{-1}$, and it is speculated these
may be in the process of merging. Such a scenario may be consistent
with the detection of widespread ($>$~pc-scale) SiO emission, a shock
tracer, by {\it Jim\'enez-Serra et al.}  (2010) along this IRDC.
However, in general it is difficult to be certain about flow
geometries from only line of sight velocity information.  While
infall/converging flow signatures have been claimed via inverse
P-Cygni profiles in star-forming cores and clumps
(\S\ref{S:accretion}), there are few such claims in starless objects
({\it Beuther et al.,} 2013a). The L1544 PSC has $\sim 8\:M_\odot$
and an infall speed of $\simeq 0.1\:{\rm km\:s^{-1}}$ on scales of
10$^3$~AU---subsonic and $\ll v_{\rm ff}$ ({\it Keto and Caselli,}
2010).

Given a measurement of cloud velocity dispersion, $\sigma$, the extent
to which it is virialized can be assessed, but with the caveat that
the amount of $B$-field support is typically unknown. Comparing
$\rm ^{13}CO$-derived $\sigma$s with MIR+NIR extinction masses, {\it
  Kainulainen and Tan} (2013) found $\bar{\alpha}_{\rm vir}=1.9$. {\it
  Hernandez et al.} (2012) compared MIR+NIR extinction masses with
$\rm C^{18}O$-derived $\sigma$s and surface pressures in
strips across IRDC G035.39–00.33, finding results consistent with
virial equilibrium ({\it Fiege and Pudritz}, 2000).

For starless cores, {\it Pillai et al.} (2011) studied the dynamics of
cold cores near UC \ion{H}{2} regions using $\rm NH_2D$-derived
$\sigma$ and 3.5~mm emission to measure mass, finding
$\bar{\alpha}_{\rm vir}\sim 0.3$. {\it Tan et al.} (2013b) measured
mass and $\Sigma$ from both MIR+NIR extinction and mm dust emission to
compare predictions of the Turbulent Core Accretion model (including
surface pressure confinement and Alfv\'en Mach number ${\cal M}_A=1$
magnetic support) with observed $\sigma$, derived from $\rm
N_2D^+$. In six cores they found a mean ratio of observed to predicted
velocity dispersions of $0.81\pm 0.13$. However, for the massive
monolithic core C1-S they found a ratio of $0.48\pm0.17$, which at
face value implies sub-virial conditions. However, virial equilibrium
could apply if the magnetic fields were stronger so that ${\cal
  M}_A\simeq 0.3$ rather than 1, requiring $B\simeq$~1.0~mG. {\it
  S\'anchez-Monge et al.}  (2013c) used \AMM-derived mass and $\sigma$
to find $\alpha_{\rm vir}\sim 10$ for several tens of mostly low-mass
starless cores, which would suggest they are unbound. However, they
also found a linear correlation of $M$ with virial mass $M_{\rm vir}
\equiv \alpha_{\rm vir} M$, only expected if cores are
self-gravitating, so further investigation of the accuracy of the
absolute values of $\alpha_{\rm vir}$ is needed.

\subsection{Chemical Properties of Starless Cores \& Clumps}\label{S:astrochem}

IRDC chemical properties resemble those of low-mass dense cores (e.g.,
{\it Vasyunina et al.,} 2012; {\it Miettinen et al.,} 2011; {\it
  Sanhueza et al.,} 2013), with widespread emission of NH$_3$ and
N$_2$H$^+$ (e.g., {\it Zhang et al.,} 2009; {\it Henshaw et al.,}
2013). In the Nobeyama survey of {\it Sakai et al.} (2008), no CCS was
detected, suggesting the gas is chemically evolved, i.e., atomic
carbon is mostly locked into CO.

\subsubsection{CO Freeze-Out}

CO is expected to freeze-out from the gas phase onto dust grains when
$T_d\lesssim 20$~K (e.g., {\it Caselli et al.,} 1999). The CO
depletion factor, $f_D({\rm CO})$, is defined as the ratio of the {\it
  expected} CO column density given a measured $\Sigma$ (assuming
standard gas phase abundances, e.g., $n_{\rm CO}/n_{\rm H_2} =
2\times10^{-4}$, {\it Lacy et al.,}~1994) to the {\it observed} CO
column density. {\it Miettinen et al.} (2011) compared CO(1-0) \&
(2-1) observations with $\Sigma$ derived from FIR/mm emission finding
no evidence for depletion. {\it Hernandez et al.} (2012) compared NIR
\& MIR-extinction-derived $\Sigma$ with $\rm C^{18}O$(2-1) \& (1-0) to
map $f_D$ in IRDC G035.39-00.33, finding widespread depletion with
$f_D \sim 3$.  {\it Fontani et al.} (2012) compared \COII(3-2) with
FIR/mm-derived $\Sigma$ in 21 IRDCs and found $\bar{f}_D \sim 30$,
perhaps due to CO(3-2) tracing higher density (shorter depletion
timescale) regions.  On the other hand, {\it Zhang et al.} (2009)
found $f_D\sim 10^2 - 10^3$ in IRDC G28.34+0.06 P1 \& P2 by comparing
$\rm C^{18}O$(2-1) to $\Sigma$ from FIR/mm emission.

\subsubsection{Deuteration}

Freeze-out of CO and other neutrals boosts the abundance of ({\it
  ortho-})H$_2$D$^+$ and thus the deuterium fractionation of other
species left in the gas phase ({\it Dalgarno and Lepp,} 1984).
Low-mass starless cores on the verge of star formation, i.e., PSCs,
show an increase in \Dfrac(\NNH) $\equiv N$(\NND)/$N$(\NNH) to
$\gtrsim 0.1$ ({\it Crapsi et al.,}~2005). High ({\it
  ortho-})H$_2$D$^+$ abundances are also seen ({\it Caselli et al.,}
2008).  In the protostellar phase, \Dfrac (\NNH) \& $N({\rm H_2D^+})$
decrease as the core envelope is heated ({\it Emprechtinger et
  al.,}~2009; {\it Ceccarelli et al.,} this volume).

To see if these results apply to the high-mass regime, Fontani et
al.~(2011) selected core/clumps, both starless and those associated
with later stages of massive star formation, finding: (1) the average
\Dfrac (\NNH ) in massive starless core/clumps located in quiescent
environments tends to be as large as in low-mass PSCs ($\sim 0.2$);
(2) the abundance of \NND\ decreases in core/clumps that either harbor
protostars or are starless but externally heated and/or shocked (see
also {\it Chen et al.,}~2011; {\it Miettinen et al.,}~2011).

$\rm HCO^+$ also becomes highly deuterated, but as CO freezes-out,
formation rates of both $\rm HCO^+$ and $\rm DCO^+$ drop, so $\rm
DCO^+$ is not such a good tracer of PSCs as \NND. Deuteration of
\AMM\ is also high ($\gtrsim 0.1$) in starless regions of IRDCs, but,
in contrast to \NND, can remain high in the protostellar phase (e.g.,
{\it Pillai et al.,}~2011), perhaps since NH$_2$D \& \AMM\ also form
in grains mantles, unlike \NNH\ \& \NND, so abundant NH$_2$D can
result from mantle evaporation.  DNC/HNC are different, with smaller
\Dfrac(HNC) in colder, earlier-stage cores ({\it Sakai et al.,} 2012).

\subsubsection{Ionization Fraction}

The ionization fraction, $x_e$, helps set the ambipolar diffusion
timescale, $t_{\rm ad}$, and thus perhaps the rate of PSC contraction.
Observations of the abundance of molecular ions like H$^{13}$CO$^+$,
DCO$^+$, \NNH\ \& \NND\ can be used to estimate $x_e$ (Dalgarno
2006). {\it Caselli et al.}~(2002) measured $x_e \sim 10^{-9}$ in the
central regions of PSC L1544, implying $t_{\rm ad} \simeq t_{\rm
  ff}$. In massive starless cores, {\it Chen et al.}~(2011) and {\it
  Miettinen et al.}~(2011) found $x_e \sim 10^{-8}-10^{-7}$,
implying either larger cosmic-ray ionization rates or lower densities
than in L1544. However, accurate estimates of $x_e$ require detailed
chemical modeling, currently lacking in the above studies, as well as
knowledge of core structure---typically not well constrained. Core
$B$-fields can also affect low-energy cosmic ray penetration,
potentially causing variation in cosmic-ray ionization rate
({\it Padovani and Galli,} 2011). 

\subsection{Effect of Cluster Environment}

The cluster environment may influence the physical and chemical
properties of PSCs due to, e.g., warmer temperatures, enhanced
turbulence, and (proto-)stellar interactions.  Surveys of cores in
cluster regions have started to investigate this issue, but have so
far mostly targeted low-mass star-forming regions, like Perseus (e.g.,
{\it Foster et al.,} 2009). These studies find cores have higher
kinetic temperatures ($\sim 15$~K) than isolated low-mass cores
($\sim10$~K). In spite of turbulent environments, cores have mostly
thermal line widths.

Studies of proto-clusters containing an intermediate- or high-mass
forming star (e.g.,~IRAS 05345+3157, {\it Fontani et al.,} 2008;
G28.34+0.06, {\it Wang et al.,}~2008; W43, {\it Beuther et al.,} 2012)
have shown starless cores can have supersonic internal motions and
\Dfrac (\AMM, \NNH) values similar to low-mass star-forming
regions. {\it S\'anchez-Monge et al.}~(2013c) analyzed {\it VLA}
\AMM\ data of 15 intermediate-/high-mass star-forming regions, finding
73 cores, classified as quiescent starless, perturbed starless or
protostellar. The quiescent starless cores have smaller line widths
and gas temperatures ($1.0\:{\rm km\:s^{-1}}$; 16~K), than perturbed
starless ($1.4\:{\rm km\:s^{-1}}$; 19~K) and protostellar ($1.8\:{\rm
  km\:s^{-1}}$; 21~K) cores. Still, even the most quiescent starless
cores possess significant non-thermal components, contrary to the
cores in low-mass star-forming regions. A correlation between core
temperature and incident flux from the most massive star in the
cluster was seen. These findings suggest the initial conditions of
star formation vary depending on the cluster environment and/or
proximity of massive stars.

\subsection{Implications for Theoretical Models}

The observed properties of PSCs, including their dependence on
environment, constrain theoretical models of (massive) star
formation. E.g., the massive ($\sim 60\:M_\odot$), cold ($T_d\sim
10$~K), highly deuterated, monolithic starless core shown in
Fig.~\ref{fig:shuo} contains many Jeans masses, has modestly
supersonic line widths, and requires relatively strong, $\sim$~mG
magnetic fields if it is in virial equilibrium. More generally, the
apparently continuous, power-law distribution of the shape of the low-
to high-mass starless CMF implies fragmentation of dense molecular gas
helps to shape the eventual stellar IMF. Improved observations of the
PSC mass function (e.g., as traced by cores showing high deuteration
of \NNH) are needed to help clarify this connection.

\section{\textbf{OBSERVATIONS OF THE ACCRETION PHASE}}\label{S:accretion}

\subsection{Clump and Core Infall Envelopes}\label{S:infall}

Infall motions can be inferred from spectral lines showing an inverse
P-Cygni profile. This results from optically thick line emission from
a collapsing cloud with a relatively smooth density distribution and
centrally-peaked excitation temperature. The profile shows emission
on the blue-shifted side of line center (from gas approaching us on
the cloud's far side) and self-absorption at line center and on the
red-shifted side. Detection of a symmetric optically thin line profile
from a rarer isotopologue helps confirm infall is being seen, rather
than just independent velocity components.

Infall to low-mass protostars is seen via spectral lines tracing
densities above $\sim$10$^4$~cm$^{-3}$ showing such inverse P-Cygni
profiles (e.g., {\it Mardones et al.,} 1997). Infall in high-mass
protostellar cores is more difficult to find, given their typically
larger distances and more crowded environments. It can also be
difficult to distinguish core from clump infall.  Single-dish
observations of HCN, CS, HCO$^+$, CO, \& isotopologues (e.g., {\it Wu
  and Evans,} 2003; {\it Wu et al.} 2005b; {\it Fuller et al.,} 2005;
{\it Barnes et al.} 2010; {\it Chen et al.,} 2010; {\it
  L\'opez-Sepulcre et al.,} 2010; {\it Schneider et al.,} 2010; {\it
  Klaassen et al.,} 2012; {\it Peretto et al.,} 2013) reveal evidence
of infall on scales $\sim 1$~pc, likely relevant to the
clump/protocluster. Derived infall velocities and rates range from
$\sim 0.2 - 1\:{\rm km\:s^{-1}}$ and $\sim 10^{-4}-10^{-1}\:M_\odot
{\rm yr}^{-1}$. However, these rates are very uncertain. {\it
  L\'opez-Sepulcre et al.} (2010) suggest they may be upper limits as
the method assumes most clump mass is infalling, whereas the
self-absorbing region may be only a lower-density outer layer.

Clump infall times, $t_{\rm infall} \equiv M_{\rm cl}/\dot{M}_{\rm
  infall}$ can be compared to $t_{\rm ff}$. E.g., {\it Barnes et al.}
(2010) measured $\dot{M}_{\rm infall} \sim 3\times
10^{-2}\:M_\odot\:{\rm yr}^{-1}$ in G286.21+0.17, with $M_{\rm cl}\sim
10^4\:M_\odot$ and $R_{\rm cl}\simeq 0.45$~pc (Fig. 1). Thus $t_{\rm
  infall}/t_{\rm ff} \simeq 3.3\times 10^5\:{\rm yr} / 5.0\times
10^4\:{\rm yr} = 6.7$. Note, this clump has the largest infall rate
out of $\sim 300$ surveyed by {\it Barnes et al.}  (2011). Similar
results hold for the $\sim 10^3\:M_\odot$ clumps NGC 2264 IRS 1 \& 2
({\it Williams and Garland,} 2002) with $t_{\rm infall}/t_{\rm ff} =
14, 8.8$, respectively.  For the central region of SDC335 studied by
{\it Peretto et al.}  (2013), with $M_{\rm cl} = 2600\:M_\odot$,
$R_{\rm cl}=0.6$~pc and $\dot{M}_{\rm infall}= 2.5\times
10^{-3}\:M_\odot\:{\rm yr}^{-1}$ (including boosting factor of 3.6 to
account for accretion outside observed filaments), then $t_{\rm
  infall}/t_{\rm ff} = 7.0$. This suggests clump/cluster assembly is
gradual, allowing establishment of approximate pressure equilibrium
({\it Tan et al.,} 2006).

On the smaller scales of protostellar cores, for bright embedded
continuum sources, infall is inferred from red-shifted line profiles
seen in absorption against the continuum (the blue-shifted inverse
P-Cygni emission profile is difficult to distinguish from the
continuum). In a few cases, this red-shifted absorption is observed in
NH$_3$ at cm wavelengths against free-free emission of an embedded HC
\ion{H}{2} region (G10.62--0.38, {\it Sollins et al.,} 2005, note {\it
  Keto,} 2002 has also reported ionized gas infall in this source
(\S\ref{S:HII}); G24.78+0.08 A1, {\it Beltr\'an et al.,} 2006).  In
other cases, it is observed with mm interferometers in CN, C$^{34}$S,
$^{13}$CO against core dust continuum emission (W51 N, {\it Zapata et
  al.,} 2008; G19.61--0.23, {\it Wu et al.,} 2009; G31.41+0.31, {\it
  Girart et al.,} 2009; NGC 7538 IRS1, {\it Beuther et al.,}
2013b). {\it Wyrowski et al.} (2012) saw absorption of rotational
NH$_3$ transitions against FIR dust emission with {\it SOFIA}.  For
the interferometric observations, infall on scales of $\sim 10^3$~AU
is traced. Infall speeds are a few km\,s$^{-1}$; $\dot{M}_{\rm
  infall}\sim 10^{-3}$--$10^{-2}$ $M_\odot$\,yr$^{-1}$. {\it Goddi et
  al.} (2011a) used $\rm CH_3OH$ masers in AFGL 5142 to infer
$\dot{M}_{\rm infall}\sim 10^{-3}\:M_\odot\:{\rm yr}^{-1}$ on 300~AU
scales.  The above results indicate collapse of cores, in contrast to
clumps, occurs rapidly, i.e., close to free-fall rates.

Dust continuum polarization is observed towards some massive
protostars to infer $B$-field orientations (e.g., {\it Tang et al.,}
2009; {\it Beuther et al.,} 2010b; {\it Sridharan et al.,} 2014). {\it
  Girart et al.,} (2009) observed a relatively ordered ``hourglass''
morphology in G31.41+0.31, suggesting contraction has pinched the
$B$-field. However, since the region studied is only moderately
supercritical ($\Sigma \sim 5\:{\rm g\:cm^{-2}}$ and plane of sky $B
\sim 2.5$~mG; {\it Frau, Girart \& Beltr\'an,} priv. comm.), the field
may still be dynamically important, e.g., in transferring angular
momentum and suppressing fragmentation.

\subsection{Accretion Disks}\label{S:disks}

\begin{figure}[nt]
\includegraphics[width=8cm]{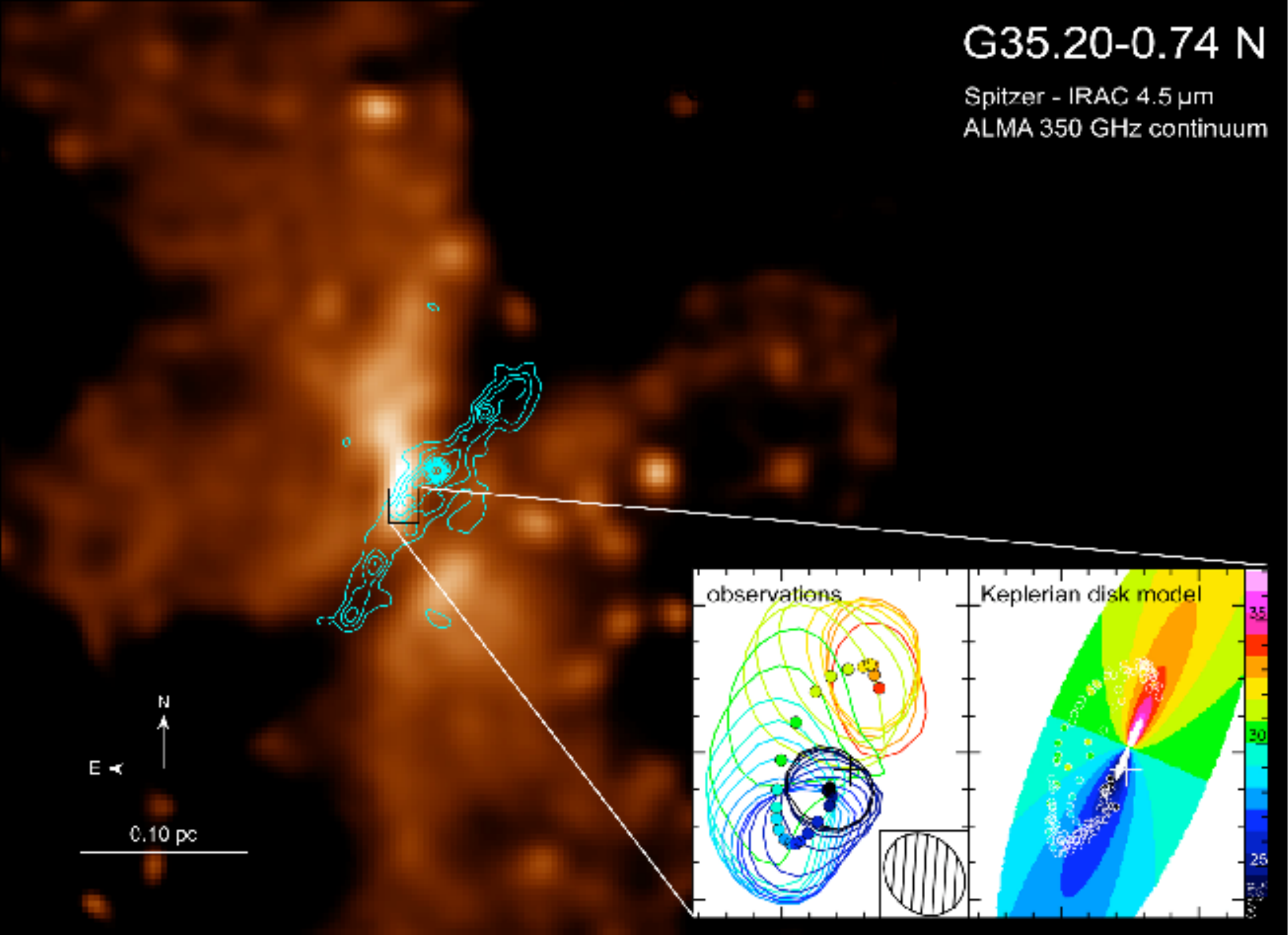}
\caption{\small 
G35.20--0.74N massive protostar ({\it S\'anchez-Monge et al.,}
2013a). Large-scale image of 4.5~$\mu$m emission, expected to trace
outflow cavities, with contours showing 850~$\rm \mu m$ continuum
observed with {\it ALMA}.
Left inset shows CH$_3$CN\,(19-18) $K$=2 emission peaks (solid
circles; outer circle is 50\% contour) from a 2D Gaussian fit
channel by channel (velocity scale on right).
Right inset compares these emission peaks with
a Keplerian model.
\label{fig:g35}
}  
 \end{figure}

In Core Accretion models, the infall envelope is expected to
transition from near radial infall to gradually greater degrees of
rotational support, until near circular orbits are achieved in a
disk. If the disk is massive, then one does not expect a Keplerian
velocity field. Also, massive moderately gravitationally unstable
disks may form large-scale, perhaps lop-sided, spiral arms that may
give the disk an asymmetric appearance ({\it Krumholz et al.,}
2007c). Disk gravitational instability is a likely mechanism to form
binaries or small-$N$ multiples. Once the infall envelope has
dispersed, either by feedback or exhaustion via accretion, then a
remnant, lower-mass, near-Keplerian disk may persist for a time, until
it also dissipates via feedback or accretion.

One of the simplest methods by which detection of accretion disks has
been claimed is via imaging of a flattened NIR-extinction structure
surrounding a NIR source (e.g., {\it Chini et al.,} 2004; {\it
  Preibisch et al.,} 2011). The latter authors report a 5500~AU
diameter disk of $2\:M_\odot$ around a $10-15\:M_\odot$ star. However,
in the absence of kinematic confirmation from molecular line
observations, one must also consider the possibility of chance
alignment of the source with a non-rotationally-supported filamentary
dust lane.

Hot and warm dust in close, $\lesssim 100$~AU, proximity to the
protostar, likely in a disk or outflow cavity wall, can sometimes be
inferred from NIR or MIR interferometry (e.g., {\it Kraus et al.,}
2010; {\it de Wit et al.,} 2011; {\it Boley et al.,} 2013). Most of
the 24 MIR sources studied by {\it Boley et al.} show deviations from
spherical symmetry, but it is difficult to tell if these are due
primarily to disks or outflows.

For methods tracing kinematics, there are $\sim 10$ examples of
``rotating toroids'' in which velocity gradients traced by, e.g.,
C$^{34}$S, HDO, H$^{18}$O, or $\rm CH_3CN$, have been seen on $\gtrsim
{\rm few} \times 1000$~AU scales that are perpendicular to
protostellar outflows (\S\ref{S:outflows}) emerging from ``hot cores''
(\S\ref{S:hotcore}) ({\it Cesaroni et al.,} 2007; {\it Beltr\'an et
  al.,} 2011). Most are probably in the process of forming B stars,
such as AFGL~2591--VLA3 ({\it Wang et al.,} 2012), but the sample also
includes the UC \ion{H}{2} regions G10.62--0.38 and G29.96--0.02 ({\it
  Beltr\'an et al.,} 2011) with $m_*\gtrsim 15\:M_\odot$. The disk
reported by {\it Wang et al.,} (2012) appears to have sub-Keplerian
kinematics, together with an expanding component perhaps driven by the
outflow. Keplerian rotation has been claimed for IRAS~20126+4104 ({\it
  Cesaroni et al.,} 2005) and G35.20--0.74N ({\it S\'anchez-Monge et
  al.,} 2013a; Fig.~\ref{fig:g35}). However, in the latter, where a
$r_d\gtrsim$2500~AU disk is inferred from an arc of centroid positions
of sequential velocity channels of $\rm CH_3CN$ observed with
$\sim$1000~AU resolution, there is misalignment of the projected
rotation axis with the N-S MIR and ionized jet, thought to define the
outflow axis ({\it Zhang et al.,} 2013b; see \S\ref{S:outflows}).  On
smaller scales, usually in nearby, lower mass and luminosity ($\sim
10^4\:L_\odot$) systems, there is also evidence of flattened
structures with kinematic gradients perpendicular to outflows (e.g.,
Cep A HW2, {\it Patel et al.,} 2005; IRAS 16547-4247, {\it
  Franco-Hern\'andez et al.,} 2009; IRAS 18162-2048, {\it
  Fern\'andez-L\'opez et al.,} 2011, {\it Carrasco-Gonz\'alez et al.,}
2012).

NIR spectroscopic observations of CO(2-0) bandhead emission, sometimes
emerging via scattered light through outflow cavities, can provide
information about protostellar and disk photospheres, where
temperatures are $\sim$2000-5000~K, i.e., scales $\lesssim$ few~AU
(e.g., {\it Bik and Thi,} 2004; {\it Davies et al.,} 2010; {\it Ilee
  et al.,} 2013). With spectral resolutions $\gtrsim 10^4$, disk
kinematics can begin to be resolved. In {\it Ilee et al.'s} study, all
20 sources can be fit with a Keplerian model.  For radio source {\it
  I} in the Orion Kleinmann-Low (KL) region (e.g., {\it Menten and
  Reid,} 1995; {\it Plambeck et al.,} 2013), photospheric temperatures
$\sim4500$~K are inferred ({\it Morino et al.,} 1998; {\it Testi et
  al.,} 2010). {\it Hosokawa and Omukai} (2009) modeled this as
emission from a very large $\sim 100\:R_\odot$ protostar (swollen by
high accretion rates), while {\it Testi et al.} (2010) preferred
accretion disk models.

There are claims of massive protostellar accretion disks based on
methanol masers (e.g., {\it Pestalozzi et al.,} 2009). However,
characterization of disks by this method is hampered by the uncertain
excitation conditions and nonlinear nature of the maser emission,
together with possible confusion with outflow motions. In Cep A HW2
methanol masers appear to trace the outflow ({\it Torstensson et al.,}
2011). Note, Zeeman splitting of these maser lines allow $B$-field
strengths ($\sim$20~mG) and morphologies (perpendicular to the disk)
to be measured ({\it Vlemmings et al.,} 2010).

Spatially and kinematically well-resolved observations via thermal
line emission from massive protostar disks remain lacking. This is not
surprising if disk diameters are typically $\lesssim$1000~AU
(\S\ref{S:protostar}), i.e., $\lesssim$ 0.5\arcsec\ at 2~kpc. The high
angular resolution to be achieved by {\it ALMA} in the coming years
should provide breakthrough capabilities in this area.

\subsection{Protostellar Outflows}\label{S:outflows}

Collimated, bipolar protostellar outflows (see also {\it Frank et
  al.,} this volume) have been observed from massive protostars,
mostly via CO and HCO$^+$, and their isotopologues, (e.g., {\it
  Beuther et al.,} 2002; {\it Wu et al.,} 2005a; {\it Garay et al.,}
2007; {\it L\'opez-Sepulcre et al.,} 2009). Correlations are seen
between outflow power, force and mass loss rate, with bolometric
luminosity over a range $L \sim $0.1--10$^6 L_\odot$. This suggests
outflows from massive protostars are driven in the same way as those
from low-mass protostars, namely via magneto-centrifugal X- or
disk-winds (momentum from radiation pressure, $\sim L/c$, is far too
small; {\it Wu et al.,} 2005a).

Based on a tentative trend inferred from several sources, {\it Beuther
  and Shepherd} (2005; see also {\it Vaidya et al.,} 2011) proposed a
scenario in which outflow collimation decreases with increasing
protostellar mass, perhaps due to the increasing influence of
quasi-spherical feedback (winds, ionization or radiation
pressure). However, such evolution is also expected in models of
disk-wind breakout from a self-gravitating core ({\it Zhang et al.,}
2014; see Fig.~\ref{fig:yichen}).

Study of SiO may help disentangle ``primary'' outflow (i.e., material
launched from the disk) from ``secondary'' outflow (i.e., swept-up
core/clump material). SiO may trace the primary outflow more directly,
since its abundance is likely enhanced for the part of the disk-wind
(and all the X-wind) launched from inside the dust destruction
radius. However, SiO may also be produced in internal shocks in the
outflow or at external shocks at the cavity walls. The single-dish
survey of {\it L\'opez-Sepulcre et al.}~(2011) found a decrease of the
SiO luminosity with increasing luminosity-to-mass ratio in massive
protostars (however, see {\it S\'anchez-Monge et al.}
2013b). Interferometric SiO observations, necessary to resolve the
structure of massive protostellar outflows, are relatively few and
mostly focused on sources with $L<10^5 L_\odot$ (e.g., AFGL~5142, {\it
  Hunter et al.,} 1999; IRAS 18264--1152, {\it Qiu et al.,} 2007; IRAS
18566+0408, {\it Zhang et al.,} 2007; G24.78+0.08, {\it Codella et
  al.,} 2013; IRAS 17233--3606, {\it Leurini et al.,} 2013). These
have traced well-collimated jets, with collimation factors similar to
those from low-mass protostars.  For higher-mass protostars with
$L>10^5\:L_\odot$, interferometric SiO observations are scarcer and
collimation results inconclusive. {\it Sollins et al.,}~(2004) mapped
the shell-like UC \ion{H}{2} region G5.89$-$0.39 with the {\it SMA},
finding a collimated SiO outflow, but the resolution was insufficient
to distinguish the multiple outflows later detected in CO by {\it Su
  et al.} (2012). On the other hand, for IRAS 23151+5912 ({\it Qiu et
  al.,} 2007) and IRAS 18360--0537 ({\it Qiu et al.,} 2012), the
molecular outflows traced by SiO are not well collimated and are
consistent with ambient gas being entrained by an underlying
wide-angle wind. Vibrationally excited SiO maser emission is thought
to trace a wide-angle bipolar disk wind on scales of 10--100~AU around
the massive protostar source {\it I} in Orion KL (e.g., {\it Greenhill
  et al.,} 2013).

Thermal (bremsstrahlung) radio jets should become prominent when the
protostar contracts towards the main sequence (i.e., for $m_*\gtrsim
15\:M_\odot$) causing its H-ionizing luminosity to increase
dramatically. Shock ionization, including at earlier stages, is also
possible ({\it Hofner et al.,} 2007). The primary outflow will be the
first gas ionized, so cm continuum and radio recombination lines
(RRLs) can trace massive protostar outflows. Elongated, sometimes
clumpy, thermal radio continuum sources are observed around massive
protostars: e.g., G35.20--0.74N ({\it Gibb et al.,} 2003); IRAS
16562-3959 ({\it Guzm\'an et al.,} 2010). Many unresolved HC
\ion{H}{2} regions (\S\ref{S:HII}) may be the central parts of ionized
outflows, since the emission measure is predicted to decline rapidly
with projected radius ({\it Tan and McKee,} 2003). Synchrotron radio
jets are seen from some massive protostars, e.g., W3($\rm H_2O$) ({\it
  Reid et al.,} 1995) and HH~80--81 ({\it Carrasco-Gonz\'alez et al.,}
2010), but why some have synchrotron emission while most others are
thermal is unclear.

Outflows also manifest themselves via cavities cleared in the
core. Cavity walls, as well as exposed disk surface layers, experience
strong radiative, and possibly shock, heating, which drive
astrochemical processes that liberate and create particular molecular
species that can serve as further diagnostics of outflows, such as
water and light hydrides (see \S\ref{S:hotcore}) and maser activity
(e.g., $\rm H_2O$, $\rm CH_3OH$) (e.g., {\it Ellingsen et al.,} 2007;
{\it Moscadelli et al.,} 2013). High-J CO lines are another important
tracer of this warm, dense gas ({\it Fuente et al.,} 2012; {\it
  Y{\i}ld{\i}z et al.,} 2013; {\it San Jos{\'e}-Garc{\'{\i}}a et al.,}
2013).  Densities and temperatures of outflowing gas in AFGL 2591 have
been measured by CO and other highly excited linear rotors ({\it van
  der Wiel et al.,} 2013).

Given the high extinctions, outflow cavities can be the main escape
channel for MIR (and even some FIR) radiation, thus affecting source
morphologies. {\it De Buizer} (2006) proposed this explains the 10 \&
20~$\rm \mu m$ appearance of G35.20--0.74N, where only the northern
outflow cavity that is inclined towards us (and is aligned with the
northern radio jet) is prominent in ground-based imaging. {\it Zhang
  et al.} (2013b) observed this source with {\it SOFIA} at
wavelengths up to $\sim$40~${\rm \mu m}$ and detected the fainter
counter jet. Comparing with Core Accretion RT models, they estimated
the protostar has $m_*\sim 20 - 34\:M_\odot$, embedded in a core with
$M_c=240\:M_\odot$, in a clump with $\Sigma_{\rm cl} \simeq 0.4 -
1\:{\rm g\:cm^{-2}}$.

While much MIR emission from outflow cavities is due to thermal
heating of cavity walls, in the NIR a larger fraction is emitted from
the protostar and inner disk/outflow, reaching us directly or via
scattering. The Br$\gamma$ line and rovibrational $\rm H_2$ lines in
the NIR can reveal information about the inner outflow (e.g., {\it
  Cesaroni et al.,} 2013). Polarization vectors from scattered light
can help localize the protostar: e.g., in Orion KL at 4~$\rm \mu m$
({\it Werner et al.,} 1983) these vectors point to a location
consistent with radio source {\it I}.

The Orion KL region also serves as an example of the rare class of
``explosive'' outflows. Forming the inner part of the outflow from KL,
is a wide-angle flow containing ``bullets'' of NIR $\rm H_2$ and Fe
line emission ({\it Allen and Burton,} 1993; {\it Bally et al.,}
2011).  Their spectra and kinematics yield a common age of $\sim
10^3$~yr. A 10$^4$~yr-old example of such a flow has been claimed by
{\it Zapata et al.} (2013) in DR21. The KL outflow has been
interpreted as being due to tidally-enhanced accretion and thus
outflow activity from the close ($\sim {\rm few}\times 10^2$~AU)
passage near source {\it I} of the Becklin-Neugebauer (BN) runaway
star, itself ejected from interaction with the $\theta^1C$ binary in
the Trapezium ({\it Tan,} 2004). BN's ejection from $\theta^1C$ has
left a distinctive dynamical fingerprint on $\theta^1C$, including
recoil motion, orbital binding energy and eccentricity---properties
unlikely (probability $\lesssim 10^{-5}$) to arise by chance ({\it
  Chatterjee and Tan,} 2012). This scenario attributes the
``explosive'' outflow as being the perturbed high activity state of a
previously normal massive protostellar outflow (akin to an FU Orionis
outburst, but triggered by an external encounter rather than an
internal disk instability). Alternatively, {\it Bally and Zinnecker}
(2005) and {\it Goddi et al.} (2011b) proposed BN was ejected by
source {\it I}, which must then be a hard binary or have suffered a
merger. This would imply much closer, disruptive dynamical
interactions involving the massive protostar(s) at source {\it I}
({\it Bally et al.,} 2011). In either scenario, recent perturbation
of gas on $\sim 10^2-10^3$~AU scales around source {\it I} has
occurred, likely affecting observed hot core complexity ({\it Beuther
  et al.,} 2006) and interpretation of maser disk and outflow
kinematics ({\it Greenhill et al.,} 2013).

\vspace{-0.01in}
\subsection{Ionized Gas}\label{S:HII}

Observationally, HC and UC \ion{H}{2} regions are defined to have
sizes $<0.01$~pc and $<0.1$~pc, respectively ({\it Beuther et al.,}
2007; {\it Hoare et al.,} 2007). They have rising radio spectral
indices, due to thermal bremsstrahlung emission from $\sim 10^4$~K
plasma. A large fraction of HC \ion{H}{2} regions show broad (FWHM
$\gtrsim 40\:{\rm km\:s^{-1}}$) RRLs ({\it Sewilo et al.,}
2011). Studies of Brackett series lines in massive protostars also
show broad lines, perhaps consistent with disk or wind kinematics
({\it Lumsden et al.,} 2012). Demographics of the UC \ion{H}{2} region
population imply a lifetime of $\sim 10^5$~yr ({\it Wood and
  Churchwell,} 1989; {\it Mottram et al.,} 2011), much longer than the
expansion time at the ionized gas sound speed, so a confinement or
replenishment mechanism is needed.

The above observational classification may mix different physical
states that are expected theoretically during massive star formation
(\S\ref{S:protostar} \& \ref{S:feedback}). An outflow-confined
\ion{H}{2} region ({\it Tan and McKee,} 2003) is expected first,
appearing as a radio jet that gradually opens up as the entire
primary-outflow-filled cavity is ionized. Together with outflow
feedback, ionization should then start to erode the core infall
envelope, driving a photoevaporative flow. Strongly-bound parts of
the core may become ionized yet continue to accrete ({\it Keto,}
2007), as inferred in G10.62--0.38 by {\it Keto} (2002) and W51e2 by
{\it Keto and Klaassen} (2008). Eventually, remnant equatorial
accretion may continue to feed a disk that is subject to
photoevaporation (\S\ref{S:feedback}).

Since massive star ionizing luminosities vary by factors $\sim 100$
from B to O stars (Fig.~\ref{fig:evolution}), \ion{H}{2} region sizes
will also vary. So while, in general, one expects earlier phases of
accretion to be associated with smaller \ion{H}{2} regions, it is
possible some UC \ion{H}{2} regions still harbor accreting massive
protostars, while some HC \ion{H}{2} regions host non-accreting,
already-formed B stars in dense clump environments.

Using the Red {\it MSX} Source (RMS) survey ({\it Lumsden et al.,}
2013) and comparing with main sequence lifetimes, {\it Mottram et al.}
(2011) derived lifetimes of radio-quiet (RQ) massive protostars
(likely the accretion phase before contraction to the ZAMS;
Fig.~\ref{fig:evolution}) and ``compact'' (including UC) \ion{H}{2}
regions as a function of source luminosity. RQ massive protostars have
lifetimes $\simeq 5\times 10^5\:{\rm yr}$ for $L\simeq 10^4\:L_\odot$,
declining to $\simeq 1\times 10^5\:{\rm yr}$ for $L\simeq
10^5\:L_\odot$. No RQ massive protostars were seen with $L\gg
10^5\:L_\odot$, consistent with Fig.~\ref{fig:evolution}: by this
luminosity most protostars should have contracted to the ZAMS and thus
become ``radio-loud'' (for $\Sigma_{\rm cl}\lesssim 3\:{\rm
  g\:cm^{-2}}$). The ``compact'' \ion{H}{2} regions have lifetimes
$\simeq 3\times 10^5$~yr (independent of $L$).  {\it Davies et al.}
(2011) extended this work to show that the data favor models in which
the accretion rate to massive protostars increases with time, as
expected in the fiducial Turbulent Core Accretion model ({\it MT03})
with $k_\rho =1.5$ and with accretion rates appropriate for
$\Sigma_{\rm cl}\sim 1$. However, their derived trend of increasing accretion
rates is also compatible with the Competitive Accretion model of {\it
  Bonnell et al.} (2001).

\subsection{Astrochemistry of Massive Protostars}\label{S:hotcore}

Massive protostars significantly affect the chemical composition of
their surroundings. Firstly, they heat gas and dust, leading to
sublimation of icy mantles formed during the cold PSC phase (e.g.,
{\it Charnley et al.,} 1992; {\it Caselli et al.,} 1993)---the hot
core phase. Secondly, they drive outflows that shock the gas enabling
some reactions with activation energies and endothermicities to
proceed (e.g., {\it Neufeld and Dalgarno,} 1989). Knowledge of
chemical processes is vital to understand the regions traced by the
various molecular lines and thus to study the structure and dynamics
of the gas surrounding the protostar, i.e., its infall envelope, disk
and outflow (e.g., {\it Favre et al.,} 2011; {\it Bisschop et al.,}
2013), fundamental to test formation theories. Unfortunately, the
chemistry in these regions is based heavily on poorly known surface
processes, so it is important to keep gathering high sensitivity and
spectral/angular resolution data to constrain astrochemical theory, as
well as laboratory data to lessen the uncertainties in the rate and
collisional coefficients required in astrochemical and RT codes.

The majority of chemical models of these early stages of massive
protostellar evolution do not take into account shocks and focus on
three main temporal phases (see review by {\it Herbst and van
  Dishoeck,} 2009): (i) cold phase ($T \sim 10$\,K), before protostar
formation, where freeze-out, surface hydrogenation/deuteration and
gas-phase ion-neutral reactions are key processes. The main
constituents of icy mantles are H$_2$O, followed by CO \& CO$_2$, and
then by CH$_3$OH, NH$_3$ \& CH$_4$ ({\it {\"O}berg et al.,} 2011), as
O, C, N \& CO are mainly hydrogenated, since H is by far the fastest
element on the surface at such low temperatures; (ii) warm-up phase,
when the protostar starts to heat the surroundings and temperatures
gradually increase from $\sim$10 to $\ga$100\,K (e.g., {\it Viti et
  al.,} 2004); (iii) hot core phase, when all mantles are sublimated
and only gas phase chemistry proceeds (e.g., {\it Brown et al.,}
1988).  The warm-up phase is thought to be critical for surface
formation of complex molecules ({\it Garrod and Herbst,} 2006). {\it
  {\"O}berg et al.} (2013) claim their observations of N- and
O-bearing organics toward high-mass star-forming region NGC~7538 IRS9
are consistent with the onset of complex chemistry at 25-30\,K. At
these dust temperatures, hydrogen atoms evaporate, while heavier
species and molecules can diffuse more quickly within the mantles and
form species more complex than in the cold phase. {\it Garrod} (2013)
showed glycine can form in ice mantles at temperatures between 40 and
120\,K and be detected in hot cores with {\it ALMA}.  {\it Aikawa et
  al.} (2012) coupled a comprehensive gas-grain chemical network with
1D hydrodynamics, also showing that complex molecules are efficiently
formed in the warm-up phase. Early complex molecule formation in the
cold phase, perhaps driven by cosmic ray induced UV photons or dust
heating ({\it Bacmann et al.,} 2012), may need to be included in the
above models.

{\it Herschel} has discovered unexpected chemistry in massive
star-forming regions. Light hydrides such as OH$^+$ and H$_2$O$^+$,
never observed before, have been detected in absorption and weak
emission toward W3 IRS5, tracing outflow cavity walls heated and
irradiated by protostellar UV radiation (e.g., {\it Benz et al.,}
2010; {\it Bruderer et al.,} 2010).

Water abundance and kinematics have been measured towards several
massive star-forming regions, with different hot cores showing a
variety of abundance levels. {\it Neill et al.} (2013) found
abundances of $\sim$6$\times$10$^{-4}$ toward Orion KL (making H$_2$O
the predominant repository of O) and a relatively large HDO/H$_2$O
ratio ($\sim$0.003), while {\it Emprechtinger et al.,} (2013) measured
10$^{-6}$ H$_2$O abundance and HDO/H$_2$O$\sim$2$\times$10$^{-4}$ in
NGC 6334 I. Thus, evaporation of icy mantles may not always be
complete in hot cores, unlike the assumption made in astrochemical
models. It also suggests that the level of deuteration is different in
the bulk of the mantle compared to the upper layers that are first
returned to the gas phase ({\it Kalv{\= a}ns and Shmed,} 2013; {\it
  Taquet et al.,} 2013), or that these two hot cores started from
slightly different initial dust temperatures, which may highly affect
water deuteration ({\it Cazaux et al.,} 2011).  Water vapour abundance
has also been found to be low in the direction of outflows
(7$\times$10$^{-7}$; {\it van der Tak et al.,} 2010), again suggesting
ice mantles are more resistant to destruction in shocks than
previously thought. The low water abundances in the outer envelopes of
massive protostars (e.g., 2$\times$10$^{-10}$ in DR21, {\it van der
  Tak et al.,} 2010; 8$\times$10$^{-8}$ in W43-MM1, {\it Herpin et
  al.,} 2012) are likely due to water being mostly in solid form.

\subsection{Comparison to Lower-Mass Protostars}\label{S:comparison}

Some continuities and similarities were already noted between high and
low-mass starless cores, including the continuous, power-law form of
the CMF (though work remains to measure the pre-stellar
CMF) and the chemical evolution of CO freeze-out and high deuteration
of certain species. Differences include massive starless cores
having larger nonthermal internal motions, though these are
also being found in lower-mass cores in high-mass star-forming
regions ({\it S\'anchez-Monge et al.,} 2013c). Massive cores may
also tend to have higher $\Sigma$s (\S\ref{S:conditions}),
whereas low-mass cores can be found with a wider range down to
lower values.

For low-mass protostars, an evolutionary sequence from PSCs to
pre-main sequence stars was defined by {\it Lada} (1991) and {\it
  Andr\'e} (1995): Class 0, I, II, and III objects based on SEDs. As
described above, an equivalent sequence for massive protostars is not
well established. Core and Competitive Accretion models predict
different amounts and geometries of dense gas and dust in the vicinity
of the protostar. Even for Core Accretion models (and also for
low-mass protostars), the SED will vary with viewing angle for the
same evolutionary stage. For a given mass accretion rate and model
for the evolution of the protostar and its surrounding disk and
envelope, the observed properties of the system can be
calculated. Examples of such models include those of {\it Robitaille
  et al.} (2006), {\it Molinari et al.}  (2008), and {\it Zhang et
  al.,} (2014; see Figs.~\ref{fig:evolution} and \ref{fig:yichen}).

For massive protostars one expects a ``radio-quiet'' phase before
contraction to the ZAMS. A growing region, at first confined to the
disk and outflow, is heated to $\gtrsim 100$~K, exhibiting hot core
chemistry. Protostellar outflows are likely to have broken out of the
core, and be gradually widening the outflow cavities. Up to this
point, the evolution of lower mass protostars is expected to be
qualitatively similar. Contraction to the ZAMS leads to greatly
increased H-ionizing luminosities and thus a ``radio-loud'' phase,
corresponding to HC or UC \ion{H}{2} regions. Hot core chemistry will
be more widespread, but there will also now be regions (perhaps
confined to the outflow and disk surfaces) that are exposed to high
FUV radiation fields. Stellar winds from the ZAMS protostar should
become much stronger than those from low-mass protostars, especially
on crossing the ``bi-stability jump'' at $T_*\simeq 21,000$~K ({\it
  Vink et al.,} 2001).

How do observed properties of lower-mass protostars compare with
massive ones? Helping address this are studies of
``intermediate-mass'' protostars with $L \sim 100 - 10^4\:L_\odot$ and
sharing some characteristics of their more massive cousins (e.g.,
clustering, creation of photo-dissociation regions). Many are closer
than 1~kpc, allowing determination of the physical and chemical
structures at similar spatial scales as in well-studied low-mass
protostars.

For disks, while there are examples inferred to be present around
massive protostars (\S\ref{S:disks}), information is lacking about
their resolved structure or even total extent, making comparison with
lower-mass examples difficult. Most massive protostellar disks appear
to have sizes $\lesssim 10^3$~AU. If disk size is related to initial
core size, then one expects $r_d\propto R_c \propto (M_c/\Sigma_{\rm
  cl})^{1/2}$, so larger disks resulting from more massive cores may
be partly counteracted if massive cores tend to be in higher
$\Sigma_{\rm cl}$ clumps. If stronger magnetization is needed to
support more massive cores, then this may lead to more efficient
magnetic braking during disk formation. Survival of remnant disks
around massive stars may be inhibited by more efficient feedback
(\S\ref{S:feedback}). 

Comparison of outflow properties was discussed in \S\ref{S:outflows},
noting the continuity of outflow force and mass loss rates with $L$.
Similar collimation factors are also seen, at least up to $L\sim
10^5\:L_\odot$.  The rare ``explosive'' outflows may affect massive
protostars more than low-mass ones, but too few examples are known to
draw definitive conclusions.

Equivalent regions exhibiting aspects of hot core chemistry, e.g.,
formation of complex organic molecules, have been found around
low-mass (e.g., IRAS 16293--2422: {\it Cazaux et al.} 2003) and
intermediate-mass (e.g., IC 1396N, {\it Fuente et al.,} 2009; NGC
7129, {\it Fuente et al.,} 2012; IRAS 22198+6336 \& AFGL 5142 [also
  discussed in \S\ref{S:infall} \& \ref{S:outflows}], {\it Palau et
  al.,} 2011) protostars. Around low-mass protostars these regions
appear richer in O-bearing molecules like methyl-formate (CH$_3$OOCH)
and poorer in the N-bearing compounds. They have high deuteration
fractions (e.g., {\it Demyk et al.,} 2010) unlike hot cores around
massive protostars. The situation in intermediate-mass protostars is
mixed: IRAS 22198+6336 and AFGL 5142 are richer in oxygenated
molecules while NGC 7129 is richer in N-species. The recent detection
of the vibrationally excited lines of CH$_3$CN and HC$_3$N in this hot
core also points to a higher gas temperature ({\it Fuente et al.,} in
prep.). {\it Palau et al.} (2011) discussed that these hot cores can
encompass two different types of regions, inner accretion disks and
outflow shocks, helping to explain the observed diversity.

Finally, the stellar IMF for $m_*\gtrsim 1\:M_\odot$ appears
well-described by a continuous and universal power law (see
\S\ref{S:cluster}), with no evidence of a break that might evidence a
change in the physical processes involved in star formation.

In summary, many aspects of the star formation process appear to
either be very similar or vary only gradually as a function of
protostellar mass. While some of these properties remain to be
explored at the highest masses, we conclude that the bulk of existing
observations support a common star formation mechanism from low to
high masses.

\subsection{Conditions for Massive Star Formation}\label{S:conditions}

Do clumps that form massive stars require a threshold $\Sigma$ or
other special properties? {\it L{\'o}pez-Sepulcre et al.} (2010) found
an increase in outflow detection rate from 56\% to 100\% when
bisecting their clump sample by a threshold of $\Sigma_{\rm
  cl}=0.3\:{\rm g\:cm^{-2}}$. This is a factor of a few smaller than
the threshold predicted by {\it Krumholz and McKee} (2008) from
protostellar heating suppression of fragmentation of massive cores
(\S\ref{S:feedback}), and thus consistent within the uncertainties in
deriving $\Sigma$. However, this clump sample contains a mixture of
IR-dark and bright objects spanning this threshold, whereas one
expects protostellar heating to be associated with IR-bright
objects. {\it Longmore et al.} (2011) estimated the low-mass stellar
population needed to be responsible for the observed temperature
structure in the fragmenting clump G8.68--0.37, concluding it is too
large compared to that allowed by the clump's bolometric
luminosity. {\it BT12} found relatively low values of $\Sigma_{\rm
  cl}\sim 0.3\:{\rm g\:cm^{-2}}$ in their IRDC sample and advocated
magnetic suppression of fragmentation. {\it Kauffmann et al.}  (2010)
found three clouds that contain massive star formation (Orion A,
G10.15--0.34, G11.11--0.12) satisfy $M(r)\geq 870 r_{\rm
  pc}^{4/3}\:M_\odot$, while several clouds not forming massive stars
do not. This empirical condition is equivalent to $\Sigma \geq 0.054
M_3^{-1/2}\:{\rm g\:cm^{-2}}$, i.e., a relatively low threshold that
may apply to the global clump even though massive stars form in higher
$\Sigma$ peaks. In summary, more work is needed to better establish if
there are minimum threshold conditions for massive star
formation. This is difficult since once one is sure a massive star is
forming, it will have altered its environment. Thus it may be more
fruitful studing the formation requirements of massive PSCs, though
there are currently very few examples (\S\ref{S:initial}).

\section{\textbf{RELATION TO STAR CLUSTER FORMATION}}\label{S:cluster}

\vspace{-0.007in}
\subsection{The Clustering of Massive Star Formation}

{\it de Wit et al.} (2005) studied Galactic field O stars, concluding
that the fraction born in isolation was low ($4\pm2\%$). {\it Bressert
  et al.} (2012) have found a small number of O stars that appear to
have formed in isolation in the 30 Dor region of the LMC, while {\it
  Selier et al.} (2011) and {\it Oey et al.} (2013) have presented
examples in the SMC. For the Galactic sample, the low fraction of
``isolated-formation'' O stars could be modeled by extrapolating a
stochastically-sampled power-law initial cluster mass function (ICMF)
down to very low masses, including ``clusters'' of single stars. Such
a model suggests that massive star formation is not more clustered
than lower-mass star formation and that the ``clustering'' of star
formation does not involve a minimum threshold of cluster mass or
density (see also {\it Bressert et al.}  2010).

The question of whether massive stars tend to form in the central
regions of clumps/clusters is difficult to answer. Observationally,
there is much evidence for the central concentration of massive stars
within clusters (e.g., {\it Hillenbrand,} 1995; {\it Qiu et al.,}
2008; {\it Kirk and Myers,} 2012; {\it Pang et al.,} 2013; {\it Lim et
  al.,} 2013).  For clusters, like the ONC, where a substantial
fraction of the initial gas clump mass has formed stars, dynamical
evolution leading to mass segregation during star cluster formation
may overwhelm any signature of primordial mass segregation (e.g., {\it
  Bonnell and Davies,} 1998; {\it Allison and Goodwin,} 2011; {\it
  Maschberger and Clarke,} 2011), especially if cluster formation
extends over many local dynamical times ({\it Tan et al.,} 2006). This
problem is even more severe in gas-free, dynamically-older systems
like NGC 3603, Westerlund 1 and the Arches.

Earlier phase studies are needed. {\it Kumar et al.}  (2006) searched
2MASS images for clusters around 217 massive protostar candidates,
finding 54. Excluding targets most affected by Galactic plane
confusion, the detection rate was $\sim 60\%$.  {\it Palau et al.}
(2013) studied 57 (mm-detected) cores in 18 ``protoclusters'', finding
quite low levels of fragmentation and relatively few associated
NIR/MIR sources.

Do massive stars tend to form earlier, later or contemporaneously with
lower-mass stars? In Turbulent Core Accretion ({\it MT03}), the
formation times of stars from their cores show a weak dependence with
mass, $t_{*f}\propto m_{*f}^{1/4}$, and the overall normalization is
short compared to the global cluster formation time, if that is spread
out over at least a few free-fall times. Competitive Accretion models
(e.g., {\it Wang et al.,} 2010) involve massive stars gaining their
mass gradually over the same timescales controlling global clump
evolution, suggesting that massive stars would form later than typical
low-mass stars.  Systematic uncertainties in young stellar age
estimates ({\it Soderblom et al.,} this volume) make this a
challenging question to answer. In the ONC, {\it Da Rio et al.} (2012)
have shown there is a real age spread of a few Myr, i.e., at least
several mean free-fall times, but no evidence for a mass-age
correlation. Massive stars are forming today in the ONC, i.e., source
{\it I}. If the runaway stars $\mu$~Col and AE Aur, together with the
resulting binary, $\iota$~Ori, were originally in the ONC about
2.5~Myr ago ({\it Hoogerwerf et al.,} 2001), then massive stars appear
to have formed contemporaneously with the bulk of the cluster
population.

\vspace{-0.01in}
\subsection{The IMF and Binarity of Massive Stars}\label{S:IMF}

The massive star IMF and its possible variation with environment are
potential tests of formation mechanisms. The IMF is constrained by
observations of massive stars in young clusters, especially ``super
star clusters'' (SSCs) with $M_{*} \gtrsim 10^4\:M_\odot$ where
effects of incomplete statistical sampling are reduced.  E.g., for IMF
$dN/dm_* \propto m_*^{-\alpha_*}$ and $\alpha_*=2.35$ (Salpeter) from
$m_{*l}=0.1\:M_\odot$ to upper truncation mass of $m_{*u} = 100$ or
$1000\:M_\odot$, the median expected maximum stellar mass in a cluster
with $M_*=10^4\:M_\odot$ is $83.6, 226\:M_\odot$,
respectively (e.g., {\it McKee and Williams,} 1997). For $M_*=10^5\:M_\odot$
it is $98.0, 692\:M_\odot$, respectively. 

Estimates of $m_{*u}$ range from $\simeq 150\:M_\odot$ (e.g., {\it
  Figer,} 2005) to $\simeq 300\:M_\odot$ ({\it Crowther et al.,}
2010), with uncertainties due to crowding, unresolved binarity,
extinction corrections and the NIR magnitude-mass relation. Also, a
limiting $m_{*u}$ arising from star formation may occasionally appear
to be breached by mergers or mass transfer in binary systems (e.g.,
{\it Banerjee et al.,} 2012; {\it Schneider et al.,} 2014). It is not
yet clear if $m_{*u}$ is set by local processes (e.g., ionization or
radiation pressure feedback (\S\ref{S:feedback}, \S\ref{S:numerical}),
rapid mass loss due to stellar instability) or by the cluster
environment (e.g., {\it Weidner et al.,} 2013; however, see {\it
  Krumholz,} 2014).

Deriving initial stellar masses can thus require modeling stellar
evolution, including the effects of rotation, mass loss and binary
mass transfer (e.g., {\it Sana et al.,} 2012; {\it De} {\it Mink et
  al.,} 2013; {\it Schneider et al.,} 2014).  Dynamical evolution in
clusters leads to mass segregation and ejection of stars, further
complicating IMF estimation from observed mass functions (MFs) of
either current or initial stellar masses.

Many attempts have been made to derive MFs in SSCs.  For Westerlund 1,
the most massive young cluster in the Galaxy, {\it Lim et al.}~(2013)
find $\alpha_* = 1.8 \pm 0.1$ within $r < 2.8$~pc over mass range $5 <
m_{*}/M_\odot < 85$, and an even shallower slope of $\alpha_* = 1.5$
if the statistically incomplete highest-mass bins are excluded.  A
similar slope of $\alpha_* = 1.9 \pm 0.15$ for $1 < m_{*}/M_\odot <
100$ is measured for proper-motion members in the central young
cluster of NGC~3603, with an even shallower slope of $\alpha_* = 1.3
\pm 0.3$ found in the cluster core for the intermediate- to high-mass
stars ($4 < m_{*}/M_\odot < 100$, {\it Pang et al.,}~2013).  For R136
in 30 Doradus in the LMC, {\it Andersen et al.}~(2009) find $\alpha_*
= 2.2 \pm 0.2$ for $1 < m_{*f}/M_\odot < 20$ and a radial coverage of
3 to 7~pc.  However, the cluster core remains poorly resolved and its
MF uncertain.  In NGC~346 in the SMC, {\it Sabbi et al.} (2008) find
$\alpha_* = 2.43 \pm 0.18$ for $0.8 < m_{*f}/M_\odot < 60$.
 
Environmental conditions of temperature, cosmic ray flux,
magnetization and orbital shear are all higher in the Galactic center,
so one may expect IMF variations (e.g., {\it Morris and Serabyn,}
1996). In the Quintuplet cluster core, $r < 0.5$pc, the
present-day MF is found to be $\alpha_* = 1.7 \pm 0.2$ for $4 <
m_{*f}/M_\odot < 48$ ({\it Hu{\ss}mann et al.,}~2012). As in
Westerlund 1 and NGC~3603, a steepening of the IMF slope with distance
from cluster center is observed with $\alpha_* = 2.1 \pm 0.2$ for
radii 1.2 to 1.8~pc, close to the expected tidal radius ({\it B.
  Hu{\ss}mann,} priv. comm.). The core of the Arches cluster, also exhibits a
relatively shallow MF, but the combined MF slope out to
the tidal radius is $\alpha_* = 2.5 \pm 0.2$ for $15 < m_{*f}/M_\odot
< 80$ ({\it Habibi et al.,}~2013).

Detailed N-body simulations have been carried out to model the Arches
Cluster (e.g., {\it Harfst et al.,}~2010). The excellent match between
the radial variation of the MF in these simulations and the observed
increase in the MF slope with radius provide strong evidence that the
steepening is caused by dynamical mass segregation alone.  By analogy,
the relatively shallow slopes observed in the central regions of all
the above young, massive clusters are likely influenced, and possibly
completely caused, by internal dynamical evolution of these clusters
on timescales as short as 1-3 Myr, within the current ages of these
clusters.

The most extreme star-forming environment resolved to date is the
young nuclear cluster surrounding the supermassive black hole
SgrA$^\ast$ in the center of the Milky Way. If the effects of
increased tidal shear and temperatures cause an increase in the Jeans
mass, it should most likely be observed in this environment.  Previous
studies suggested a slope as shallow as $\alpha_* = 0.45 \pm 0.3$ for
$m_{*} > 10\,M_\odot$ and with a truncation of $m_{*u}\simeq
30\:M_\odot$, and hence proposed the most extreme stellar MF observed
in a resolved population to date ({\it Bartko et al.,}~2010).  Many of
the young stars in the nuclear cluster are in an elongated disk-like
structure (e.g., {\it Paumard et al.,}~2006), and optimizing for the
inclusion of young disk candidates as members of the cluster revises
this picture. Employing {\it Keck} spectroscopy along the known disk
of young stars, {\it Lu et al.}~(2013) found a slope of $\alpha_* =
1.7 \pm 0.2$ from detailed Bayesian modeling to derive individual
stellar masses.  While still flatter than the Salpeter slope, this
result is now in agreement with the shape of the MFs found in the
central regions of all other young, massive clusters outside of this
very extreme environment. The effects of mass segregation and ejection
for altering the observed MF are not very well known. Modulo these
uncertainties, there is no evidence for IMF variation in the Galactic
center compared to other massive clusters.

In summary, the massive young clusters resolved to date exhibit
somewhat shallow present-day MFs in their cluster cores, with a
steepening of the MF observed towards larger radii. Numerical
simulations suggest that the central top-heavy mass distribution can
be explained by mass segregation, and is not evidence for a deviating
IMF in the high-mass regime. The fact that there is little or no
variation of the shape of the high-mass IMF from NGC 346 to the Arches
or Westerlund 1 suggests that the process of massive star formation
has a very weak dependence on density, which varies by two to three
orders of magnitude between these clusters
(Fig.~\ref{fig:overview}). This implies stellar collisions are not
important for forming massive stars in these environments, in
agreement with theoretical estimates of collision rates by, e.g., {\it
  Moeckel and Clarke} (2011). Predictions of the dependence of the IMF
with density are needed from simulations and models of Core and
Competitive Accretion.

The binary properties of massive stars have been discussed extensively
by {\it Zinnecker and Yorke} (2007) (see also {\it Sana et al.,} 2012;
{\it De Mink et al.,} 2013). They are more likely to be in binary or
multiples than lower-mass stars. For stars in cluster centers, these
properties may also have been affected by dynamical evolution via
interactions (e.g., {\it Parker et al.} 2011; {\it Allison and
  Goodwin,} 2011), which tend to harden and increase the eccentricity
of binary orbits and can also lead to ejection of runaway stars.  For
example, such an interaction has been proposed to explain the
properties of the $\theta^1$~C binary near the center of the ONC ({\it
  Chatterjee and Tan,} 2012). Thus, one should be cautious using the
observed binary properties of massive stars to constrain massive star
formation theories, with attention to be focussed on objects that are
either very young (i.e., still forming) or relatively isolated in
lower-density environments.

\section{\textbf{CONCLUSIONS AND FUTURE OUTLOOK}}\label{S:conclusions}

It is a challenge to understand the wide variety of interlocking
physical and chemical processes involved in massive star
formation. Still, significant theoretical progress is being made in
modeling these processes, both individually and in combination in
numerical simulations. However, such simulations still face great
challenges in being able to adequately resolve the scales and
processes that may be important, including MHD-driven outflows,
radiative feedback and astrochemistry. There are also uncertainties in
how to initialize these simulations. Accurate prediction of the IMF,
including massive stars, of a cluster forming under given
environmental conditions remains a distant goal.

Close interaction with observational constraints is essential. Here
rapid progress is also being made and, with the advent of {\it ALMA},
this should only accelerate. One challenge is development of the
astrochemical sophistication needed to decipher the rich variety of
diagnostic tracers becoming available for both pre- and protostellar
phases. Determination of pre-stellar core mass functions and
resolution of massive protostellar accretion disks, including possible
binary formation, and outflows are important goals.

Core and Competitive Accretion theories are being tested by both
simulation and observation. Core Accretion faces challenges of
understanding fragmentation properties of magnetized, turbulent
gas, following development of accretion disks and outflows from
collapsing cores, and assessing the importance of external
interactions in crowded cluster environments. Competitive Accretion is
also challenged by theoretical implementation of realistic feedback
from MHD outflows and by observations of massive starless cores,
together with apparent continuities and similarities of the
star formation process across protostellar mass and luminosity
distributions. There is much work to be done!

\textbf{Acknowledgments.} We thank S. Kong, A. Myers,
\'A. S\'anchez-Monge, Y. Zhang for figures and discussions, and
H. Beuther, M. Butler, P. Clark, N. Da Rio, M. Hoare, P. Klaassen, P. Kroupa, W. Lim \&
a referee for discussions.

\section*{\textbf{REFERENCES}}
{\small
\baselineskip=10pt

\refs Aikawa Y., Wakelam V. et al. (2012) {\it Astrophys. J., 760}, 40.

\refs Allen D. A. and Burton M. G. (1993) {\it Nature, 363}, 54.

\refs Allison R. J. and Goodwin S. P. (2011) {\it Mon. Not. R. Astron. Soc., 415}, 1967.

\refs Andersen M., Zinnecker H. et al. (2009) {\it Astrophys. J., 707}, 1347.

\refs Andr\'e P. (1995) {\it Astrophysics and Space Science, 224}, 29.





\refs Bacmann A., Taquet V. et al. (2012) {\it Astron. Astrophys., 541}, L12.

\refs Bally J., Cunningham N. J. et al. (2011) {\it Astrophys. J., 727}, 113.


\refs Bally J. and Zinnecker H. (2005) {\it Astron. J., 129}, 2281.

\refs Banerjee S. et al. (2012) {\it Mon. Not. R. Astron. Soc., 426}, 1416.

\refs Barnes P. J. et al. (2011) {\it Astrophys. J. Suppl., 196}, 12.

\refs Barnes P. J. et al. (2010) {\it Mon. Not. R. Astron. Soc., 402}, 73.

\refs Bartko H., Martins F. et al. (2010) {\it Astrophys. J., 708}, 834.

\refs Battersby C., Bally J. et al. (2011) {\it Astron. Astrophys., 535,} 128.

\refs Bate M. R. (2012) {\it Mon. Not. R. Astron. Soc., 419}, 3115.

\refs Bate M. R. et al. (1995) {\it Mon. Not. R. Astron. Soc., 277}, 362.

\refs Baumgardt H. and Klessen R. S. (2011) {\it Mon. Not. R. Astron. Soc., 413}, 1810.





\refs Beltr\'an M. T., Cesaroni R. et al. (2006) {\it Nature, 443}, 427.


\refs Beltr\'an M. T. et al. (2011) {\it Astron. Astrophys., 525}, 151.


\refs Benz A.~O. et al. (2010) {\it Astron. Astrophys., 521}, L35.


\refs Berger M. J. and Oliger J. (1984) {\it J. Comp. Phys., 53}, 484.

\refs Bergin E. A. and Tafalla M. (2007) {\it Ann. Rev. Astron. Astrophys., 45}, 339.

\refs Bertoldi F. and McKee C. F. (1992) {\it Astrophys. J., 395}, 140.

\refs Beuther H., Schilke P. et al. (2002) {\it Astron. Astrophys., 383}, 892.

\refs Beuther H. and Schilke P. (2004) {\it Science, 303}, 1167.


\refs Beuther H. and Shepherd D. S. (2005) {\it Cores to Clusters} (M. Kumar et al.),
Springer, New York, p105-119.

\refs Beuther H., Zhang Q. et al. (2006) {\it Astrophys. J., 636}, 323.

\refs Beuther H., Churchwell E. B., Mckee C. F. and Tan J. C. (2007) {\it PPV} (B. Reipurth et al.), Univ. Arizona, Tucson, p165-180.

\refs Beuther H. et al. (2010a) {\it Astron. Astrophys., 518}, L78.

\refs Beuther H. et al. (2010b) {\it Astrophys. J., 724}, L113.

\refs Beuther H. et al. (2012) {\it Astron. Astrophys., 538}, 11.

\refs Beuther H., Linz H. et al. (2013a) {\it Astron. Astrophys., 553}, 115.

\refs Beuther H., Linz H. et al. (2013b) {\it Astron. Astrophys., 558}, 81.



\refs Bik A. and Thi W. F. (2004) {\it Astron. Astrophys., 427}, L13.

\refs Bisbas T. G. et al. (2009) {\it Astron. Astrophys., 497}, 649.

\refs Bisschop S.~E. et al. (2013) {\it Astron. Astrophys., 552}, 122.


\refs Blandford R. D. and Payne D. G. (1982) {\it Mon. Not. R. Astron. Soc., 199}, 883.

\refs Boley P. A., Linz H. et al. (2013) {\it Astron. Astrophys., 558}, 24.

\refs Bonnell I. A. and Bate M. R. (2005) {\it Mon. Not. R. Astron. Soc., 362}, 915.

\refs Bonnell I. A. et al. (2001), {\it Mon. Not. R. Astron. Soc., 323}, 785.


\refs Bonnell I. A. et al. (1998) {\it Mon. Not. R. Astron. Soc., 298}, 93.

\refs Bonnell I. A. and Davies M. B. (1998) {\it Mon. Not. R. Astron. Soc., 295}, 691.

\refs Bonnell I. A. et al. (2004) {\it Mon. Not. R. Astron. Soc., 349}, 735.

\refs Bonnell I. A., Larson R. B. \& Zinnecker H. (2007) {\it PPV} (B. Reipurth et al.), Univ. Arizona, Tucson, p149-164.

\refs Bontemps S., Motte F. et al.~(2010) {\it Astron. Astrophys., 524,} 18.




\refs Bressert E. et al. (2012) {\it Astron. Astrophys., 542}, 49.

\refs Bressert E. et al. (2010) {\it Mon. Not. R. Astron. Soc., 409}, L54.

\refs Bromm V. (2013) {\it Rep. Prog. Phys., 76}, 11.

\refs Brown P.~D. et al. (1988) {\it Mon. Not. R. Astron. Soc., 231}, 409.

\refs Bruderer S. et al. (2010) {\it Astron. Astrophys., 521}, L44.


\refs Butler M. J. and Tan J. C. (2009) {\it Astrophys. J., 696}, 484.

\refs Butler M. J. and Tan J. C. (2012) {\it Astrophys. J., 754}, 5. [BT12]



\refs Carrasco-Gonz{\'a}lez C. et al. (2010) {\it Science, 330}, 1209.

\refs Carrasco-Gonz{\'a}lez C. et al. (2012), {\it Astrophys. J., 752}, L29.

\refs Caselli P. and Myers P. C. (1995) {\it Astrophys. J., 446}, 665.

\refs Caselli P., Hasegawa T. et al. (1993) {\it Astrophys. J., 408}, 548.

\refs Caselli P., Vastel C.~et al. (2008) {\it Astron. Astrophys., 492}, 703.

\refs Caselli P., Walmsley C. M. et al. (1999) {\it Astrophys. J., 523}, L165.

\refs Caselli P., Walmsley C. M. et al. (2002) {\it Astrophys. J., 565}, 344.

\refs Cazaux S., Caselli P. et al. (2011) {\it Astrophys. J., 741}, L34.

\refs Cazaux S., Tielens A. et al. (2003) {\it Astrophys. J., 593}, L51.



\refs Cesaroni R., Galli D., Lodato G. et al. (2007) {\it Protostars and Planets V} (B. Reipurth et al., eds.), p197, Univ. of Arizona, Tucson.

\refs Cesaroni R., Massi F. et al. (2013) {\it Astron. Astrophys., 549}, 146.

\refs Cesaroni R., Neri R. et al. (2005) {\it Astron. Astrophys., 434}, 1039.


\refs Charnley S.~B. et al. (1992) {\it Astrophys. J., 399}, L71.

\refs Chatterjee S. and Tan J. C. (2012) {\it Astrophys. J., 754}, 152.

\refs Chen H.-R., Liu S.-Y., Su Y. et al. (2011) {\it Astrophys. J., 743}, 196.

\refs Chen X., Shen Z.-Q., Li Ji.-J. et al. (2010) {\it Astrophys. J., 710}, 150.

\refs Chini R., Hoffmeister V. et al. (2004) {\it Nature, 429}, 155.

\refs Chira R.-A. et al. (2013) {\it Astron. Astrophys., 552}, 40.

\refs Churchwell E., Babler B., Meade M. et al. (2009) {\it PASP, 121}, 213.

\refs Clark P. C. et al. (2007) {\it Mon. Not. R. Astron. Soc., 379}, 57.

\refs Codella C. et al. (2013) {\it Astron. Astrophys., 550}, 81.

\refs Commer{\c c}on B. et al. (2011a) {\it Astrophys. J., 742}, L9.

\refs Commer{\c c}on B. et al. (2011b) {\it Astron. Astrophys., 529}, 35.

\refs Crapsi A., Caselli P. et al. (2005) {\it Astrophys. J., 619}, 379.

\refs Crowther P. A. et al. (2010) {\it Mon. Not. R. Astron. Soc., 408}, 731.

\refs Crutcher R. M. (2012) {\it Annu. Rev. Astron. Astrophys., 50}, 29.

\refs Cunningham A. J. et al. (2011) {\it Astrophys. J., 740}, 107.

\refs Da Rio N., Robberto M. et al. (2012) {\it Astrophys. J., 748}, 14.

\refs Dale J. E. et al. (2005) {\it Mon. Not. R. Astron. Soc., 358}, 291.

\refs Dalgarno A. (2006) {\it Proc. Nat. Acad. of Sci., 103}, 12269.

\refs Dalgarno A. and Lepp S. (1984) {\it Astrophys. J., 287}, L47.

\refs Davies B. et al. (2011) {\it Mon. Not. R. Astron. Soc., 416}, 972.

\refs Davies B. et al. (2010) {\it Mon. Not. R. Astron. Soc., 402}, 1504.


\refs De Buizer J. M. (2006) {\it Astrophys. J., 642}, L57.

\refs de Gouveia Dal Pino E. et al. (2012) {\it Physica Scripta, 86}, 018401.

\refs de Mink S. E., Langer N. et al. (2013) {\it Astrophys. J., 764}, 166.

\refs de Wit W. J. et al. (2011) {\it Astron. Astrophys., 526}, L5.

\refs de Wit W. J., Testi L. et al. (2005) {\it Astron. Astrophys., 437}, 247.


\refs Demyk K., Bottinelli S. et al. (2010) {\it Astron. Astrophys., 517}, 17. 

\refs Dobbs C. L. et al. (2005) {\it Mon. Not. R. Astron. Soc., 360}, 2.

\refs Draine B. T. (2011) Physics of the Interstellar and Intergalactic Medium (Princeton: Princeton Univ. Press)


\refs Duarte-Cabral A. et al. (2013) {\it Astron. Astrophys., 558}, 125.

\refs Duffin D. F. and Pudritz R. E. (2008) {\it Mon. Not. R. Astron. Soc., 391}, 1659.


\refs Edgar R. and Clarke C. (2004) {\it Mon. Not. R. Astron. Soc., 349}, 678.

\refs Ellingsen S. P., Voronkov M. A. et al. (2007) {\it IAU Symp., 242}, 213.

\refs Emprechtinger M. et al. (2009) {\it Astron. Astrophys., 493}, 89. 

\refs Emprechtinger M., Lis D.~C. et al. (2013) {\it Astrophys. J., 765}, 61.

\refs Falgarone E. et al. (2008) {\it Astron. Astrophys., 487}, 247.

\refs Fatuzzo M. and Adams F. C. (2002) {\it Astrophys. J., 570}, 210.

\refs Favre C., Despois D. et al. (2011) {\it Astron. Astrophys., 532}, 32. 


\refs Fern\'andez-L\'opez M., Girart J. M. et al. (2011) {\it Astron. J., 142}, 97.

\refs Fiege J. D. and Pudritz R. E. (2000) {\it Mon. Not. R. Astron. Soc., 311}, 85.

\refs Figer D. F. (2005) {\it Nature, 434}, 192.

\refs Fontani F., Caselli P. et al. (2008) {\it Astron. Astrophys., 477}, L45.


\refs Fontani F. et al. (2012) {\it Mon. Not. R. Astron. Soc., 423}, 2342.

\refs Fontani F., Palau A. et al. (2011) {\it Astron. Astrophys., 529}, L7.


\refs Foster J. B. et al. (2009) {\it Astrophys. J., 696}, 298.

\refs Franco-Hern{\'a}ndez R. et al. (2009) {\it Astrophys. J., 701}, 974.




\refs Fryxell B., Olson K. et al. (2000) {\it Astrophys. J. Suppl., 131}, 273.

\refs Fuente A. et al. (2009) {\it Astron. Astrophys., 507}, 1475.

\refs Fuente A., Caselli P. et al. (2012) {\it Astron. Astrophys., 540}, 75. 



\refs Fuller G. A. et al. (2005) {\it Astron. Astrophys., 442}, 949.

\refs Garay G., Mardones D. et al. (2007) {\it Astron. Astrophys., 463}, 217.

\refs Garrod R.~T. (2013) {\it Astrophys. J., 765}, 60.

\refs Garrod R.~T. and Herbst E. (2006) {\it Astron. Astrophys., 457}, 927. 



\refs Gibb A. G. et al. (2003) {\it Mon. Not. R. Astron. Soc., 339}, 198.


\refs Ginsburg A., Bressert E. et al. (2012) {\it Astrophys. J., 758}, L29.

\refs Girart J., Beltr\'an M., Zhang Q. et al. (2009) {\it Science, 324}, 1408.



\refs Girichidis P. et al. (2012) {\it Mon. Not. R. Astron. Soc., 420}, 613.

\refs Goddi C. et al. (2011a) {\it Astron. Astrophys., 535}, L8.

\refs Goddi C., Humphreys E. et al. (2011b) {\it Astrophys. J., 728}, 15.


\refs Greenhill L. J., Goddi C. et al. (2013) {\it Astrophys. J., 770}, L32.




\refs Guzm\'an A. E., Garay G. et al. (2010) {\it Astrophys. J., 725}, 734.

\refs Habibi M., Stolte A. et al. (2013) {\it Astron. Astrophys., 556}, 26.

\refs Haemmerl\'e L. et al. (2013) {\it Astron. Astrophys., 557}, 112.

\refs Harfst S. et al. (2010) {\it Mon. Not. R. Astron. Soc., 409}, 628.


\refs Hennebelle P. et al. (2011) {\it Astron. Astrophys., 528}, 72.

\refs Henning Th., Linz H. et al.~(2010) {\it Astron. Astrophys., 518,} L95.

\refs Henshaw J. et al. (2013) {\it Mon. Not. R. Astron. Soc., 428}, 3425.

\refs Herbst E. and van Dishoeck E.~F. (2009) {\it Ann. Rev. Astron. Astrophys., 47}, 427.


\refs Hernandez A. K., Tan J. C. et al. (2012) {\it Astrophys. J., 756}, L13.

\refs Herpin F., Chavarr{\'{\i}}a L., et al. (2012) {\it Astron. Astrophys., 542}, 76. 


\refs Heyer M. H., Krawczyk C. et al. (2009) {\it Astrophys. J., 699}, 1092.

\refs Hillenbrand L.~A., Meyer M.~R. et al. (1995) {\it Astron. J., 109}, 280.


\refs Hoare M. G., Kurtz S. E., Lizano S. et al. (2007) {\it PPV} (B. Reipurth et al.), Univ. Arizona, Tucson, p181-196.

\refs Hofner P., Cesaroni R. et al. (2007) {\it Astron. Astrophys., 465}, 197.

\refs Hollenbach D., Johnstone D. et al. (1994) {\it Astrophys. J., 428}, 654.

\refs Hoogerwerf R. et al. (2001) {\it Astron. Astrophys., 365}, 49.

\refs Hosokawa T. and Omukai K. (2009) {\it Astrophys. J., 691}, 823.

\refs Hosokawa T., Omukai K. et al. (2011) {\it Science, 334}, 1250.

\refs Hosokawa T., Yorke H. W. et al. (2010) {\it Astrophys. J., 721}, 478.

\refs Houde M., Phillips, T., Vaillancourt J. et al. (2009) {\it Sub-mm Astrophysics \& Tech.} (D. Lis et al.), {\it ASP Conf. Ser., 417}, 265.

\refs Hunter C. (1977) {\it Astrophys. J., 218}, 834.

\refs Hunter T., Testi L., Zhang Q. et al. (1999) {\it Astrophys. J., 118}, 477.


\refs Hu{\ss}mann B., Stolte A. et al. (2012) {\it Astron. Astrophys., 540}, 57.

\refs Ilee J. D. et al. (2013) {\it Mon. Not. R. Astron. Soc.,  429}, 2960.


\refs Jiang Y.-F., Davis S. W. et al. (2013) {\it Astrophys. J., 763}, 102.


\refs Jijina J. and Adams F. C. (1996) {\it Astrophys. J., 462}, 874.

\refs Jim\'enez-Serra I., Caselli P., Tan J. C. et al. (2010) {\it Mon. Not. R. Astron. Soc., 406}, 187.

\refs Johnston K. G. et al. (2011) {\it Mon. Not. R. Astron. Soc., 415}, 2952.



\refs Kahn F. D. (1974) {\it Astron. Astrophys., 37}, 149.

\refs Kainulainen J. and Tan J. C. (2013) {\it Astron. Astrophys., 549}, 53.

\refs Kalv{\= a}ns J. and Shmeld I. (2013) {\it Astron. Astrophys., 554}, 111.

\refs Kauffmann J., Pillai T. et al. (2010) {\it Astrophys. J., 716}, 433.


\refs Keto E. (2002) {\it Astrophys. J., 568}, 754.

\refs Keto E. (2007) {\it Astrophys. J., 666}, 976.

\refs Keto E. and Caselli P. (2010) {\it Mon. Not. R. Astron. Soc., 402}, 1625.

\refs Keto E. and Klaassen P. D. (2008) {\it Astrophys. J., 678}, L109.

\refs Kirk H. and Myers P. C. (2012) {\it Astrophys. J., 745}, 131.

\refs Klaassen P. D., Testi L. et al. (2012) {\it Astron. Astrophys., 538}, 140.

\refs K\"onigl A. and Pudritz R.~E. (2000) {\it Protostars and Planets IV}, 759.

\refs Kratter K. M. et al. (2010) {\it Astrophys. J., 708}, 1585.

\refs Kraus S., Hofmann K., Menten K. et al. (2010) {\it Nature, 466}, 339.



\refs Krumholz M. R. (2014) {\it Physics Rep.,} in press, arXiv1402.0867.

\refs Krumholz M. R., Klein R. I. et al. (2007a) {\it Astrophys. J., 656}, 959.

\refs Krumholz M. R., Klein R. I. et al. (2007c) {\it Astrophys. J., 665}, 478.


\refs Krumholz M. R., Klein R. I. et al. (2012) {\it Astrophys. J., 754}, 71.

\refs Krumholz M. R., Klein R. I. et al. (2007b) {\it Astrophys. J., 667}, 626.

\refs Krumholz M. R., Klein R. I. et al. (2009) {\it Science, 323}, 754.

\refs Krumholz M. R. and McKee C. F. (2008) {\it Nature, 451}, 1082.

\refs Krumholz M. R., McKee C. et al. (2004) {\it Astrophys. J., 611}, 399.

\refs Krumholz M. R., McKee C. F. et al. (2005a) {\it Nature, 438}, 332.

\refs Krumholz M. R., McKee C. et al. (2005b) {\it Astrophys. J., 618}, L33.

\refs Krumholz M. R. and Tan J. C. (2007) {\it Astrophys. J., 654}, 304.

\refs Kuiper R., Klahr H. et al. (2010a) {\it Astrophys. J., 722}, 1556.

\refs Kuiper R., Klahr H. et al. (2011) {\it Astrophys. J., 732}, 20.

\refs Kuiper R., Klahr H. et al. (2012) {\it Astron. Astrophys., 537}, 122.

\refs Kuiper R., Klahr H. et al. (2010b) {\it Astron. Astrophys., 511}, 81.

\refs Kuiper R. and Yorke H. W. (2013) {\it Astrophys. J., 763}, 104.

\refs Kumar M. S. N. et al. (2006) {\it Astron. Astrophys., 449}, 1033.

\refs Kunz M. W. and Mouschovias T. Ch. (2009) {\it Mon. Not. R. Astron. Soc., 399}, L94.


\refs Lacy J. H., Knacke R. et al. (1994) {\it Astrophys. J., 428}, L69.

\refs Lada C. J. (1991) {\it The Physics of Star Formation and Early Stellar Evolution} (C. Lada and N. Kylafis), Kluwer, Dordrecht, p329.

\refs Larsen S. S. et al. (2008) {\it Mon. Not. R. Astron. Soc., 383}, 263.

\refs Larson R. B. (1985) {\it Mon. Not. R. Astron. Soc., 214}, 379.

\refs Larson R. B. and Starrfield S. (1971) {\it Astron. Astrophys., 13}, 190.

\refs Leurini S., Codella C. et al. (2013) {\it Astron. Astrophys., 554}, 35.

\refs Li P. S., Martin D., Klein R. et al. (2012a) {\it Astrophys. J., 745}, 139.

\refs Li P. S., McKee C. F., and Klein R. (2012b) {\it Astrophys. J., 744}, 73.

\refs Lim B., Chun M-Y., Sung H. et al. (2013) {\it Astron. J., 145}, 46.

\refs Longmore S. N., Pillai T. et al. (2011) {\it Astrophys. J., 726}, 97.

\refs Longmore S. N. et al. (2012) {\it Astrophys. J., 746}, 117. 

\refs L\'opez-Sepulcre A. et al. (2010) {\it Astron. Astrophys., 517}, 66.

\refs L\'opez-Sepulcre A. et al. (2009) {\it Astron. Astrophys., 499}, 811.

\refs L\'opez-Sepulcre A. et al. (2011) {\it Astron. Astrophys., 526}, L2.

\refs Lu J. R., Do T., Ghez A. M. et al. (2013) {\it Astrophys. J., 764}, 155. 

\refs Lucy L. B. (1977) {\it Astron. J., 82}, 1013.


\refs Lumsden S. L. et al. (2012) {\it Mon. Not. R. Astron. Soc., 424}, 1088.

\refs Lumsden S. L. et al. (2013) {\it Astrophys. J. Suppl., 208}, 11.

\refs Mardones D., Myers P. C. et al. (1997) {\it Astrophys. J., 489}, 719.

\refs Maschberger Th. and Clarke C. J. (2011) {\it Mon. Not. R. Astron. Soc., 416}, 541.



\refs Matthews B. C. et al. (2009) {\it Astrophys. J. Suppl., 182}, 143.

\refs Matzner C. D. (2007) {\it Astrophys. J., 659}, 1394.

\refs Matzner C. D. and McKee C. F. (1999) {\it Astrophys. J., 526}, L109.

\refs Matzner C. D. and McKee C. F. (2000) {\it Astrophys. J., 545}, 364.


\refs McCrady N. and Graham J. R. (2007) {\it Astrophys. J., 663}, 844.

\refs McKee C. F. (1989) {\it Astrophys. J., 345}, 782.

\refs McKee C. F. and Hollimann, J. H. (1999) {\it Astrophys. J., 522}, 313.

\refs McKee C. F. and Ostriker E. C. (2007) {\it Ann. Rev. Astron. Astrophys. 45}, 565.

\refs McKee C. F., Li P. and Klein R. (2010), {\it Astrophys. J., 720}, 1612.

\refs McKee C. F. and Tan J. C. (2002) {\it Nature, 416}, 59.

\refs McKee C. F. and Tan J. C. (2003) {\it Astrophys. J., 585}, 850. [MT03]

\refs McKee C. F. and Tan J. C. (2008) {\it Astrophys. J., 681}, 771.

\refs McKee C. F. and Williams J. P. (1997) {\it Astrophys. J., 476}, 144.

\refs Menten K. M. and Reid M. J. (1995) {\it Astrophys. J., 445}, L157.


\refs Miettinen O. et al. (2011) {\it Astron. Astrophys., 534}, 134.

\refs Moeckel N. and Clarke C. J. (2011) {\it Mon. Not. R. Astron. Soc., 410}, 2799.


\refs Molinari S., Pezzuto S. et al. (2008) {\it Astron. Astrophys., 481}, 345.

\refs Morino J.-I., Yamashita T. et al. (1998) {\it Nature, 393}, 340.

\refs Morris M. and Serabyn E. (1996) {\it Ann. Rev. Astron. Astrophys. 34}, 645.

\refs Moscadelli L., Li J.~J. et al. (2013) {\it Astron. Astrophys., 549}, 122.


\refs Mottram J. C., Hoare M. G. et al. (2011) {\it Astrophys. J., 730}, L33.

\refs Mouschovias T. Ch. (1987) {\it Physical Processes in Interstellar Clouds} (G. Morfill \& M. Scholer), Dordrecht, Reidel, 453.


\refs Mueller K. E. et al. (2002), {\it Astrophys. J. Suppl., 143}, 469.

\refs Myers A. T., McKee C. F. et al. (2013) {\it Astrophys. J., 766}, 97.

\refs Myers P. C. and Fuller G. A. (1992) {\it Astrophys. J., 396}, 631.

\refs Najita J. R. and Shu F. H. (1994) {\it Astrophys. J., 429}, 808.

\refs Nakano T. (1989) {\it Astrophys. J., 345}, 464.


\refs Neill J., Wang S., Bergin E. et al. (2013) {\it Astrophys. J., 770}, 142.

\refs Neufeld D.~A. and Dalgarno A. (1989) {\it Astrophys. J., 340}, 869.

\refs {\"O}berg K.~I., Boogert A. et al. (2011) {\it Astrophys. J., 740}, 109.

\refs {\"O}berg K.~I., Boamah M. et al. (2013) {\it Astrophys. J., 771}, 95.



\refs Oey M. S., Lamb J. B. et al. (2013) {\it Astrophys. J.}, 768, 66.

\refs Ossenkopf V. and Henning T. (1994) {\it Astron. Astrophys., 291,} 943.


\refs Padoan P. and Nordlund A. (2007) {\it Astrophys. J., 661}, 972.

\refs Padovani M. and Galli D. (2011) {\it Astron. Astrophys., 530}, 109.



\refs Palau A., Fuente A., Girart J. et al. (2011) {\it Astrophys. J., 743}, L32.

\refs Palau A., Fuente A., Girart J. et al. (2013) {\it Astrophys. J., 762}, 120.

\refs Palla F. and Stahler S. W. (1991) {\it Astrophys. J., 375}, 288.

\refs Pang X., Grebel E., Allison R. et al. (2013) {\it Astrophys. J., 764}, 73.



\refs Parker R.~J. et al. (2011) {\it Mon. Not. R. Astron. Soc., 418}, 2565.

\refs Patel N.~A., Curiel S., Sridharan T. et al. (2005) {\it Nature, 437}, 109.

\refs Paumard T., Genzel R. et al. (2006) {\it Astrophys. J., 643}, 1011.

\refs Peretto N. and Fuller G. A. (2009) {\it Astron. Astrophys., 505}, 405.


\refs Peretto N., Fuller G. et al. (2013), {\it Astron. Astrophys., 555}, 112.

\refs Peretto N., Fuller G. et al. (2010) {\it Astron. Astrophys., 518}, L98.

\refs Pestalozzi M. R. et al. (2009) {\it Astron. Astrophys., 501}, 999.


\refs Peters T., Banerjee R. et al. (2010a) {\it Astrophys. J., 711}, 1017.

\refs Peters T., Banerjee R. et al. (2011) {\it Astrophys. J., 729}, 72.

\refs Peters T., Klessen R. S. et al. (2010b) {\it Astrophys. J., 725}, 134.

\refs Pillai T., Kauffmann J. et al. (2011) {\it Astron. Astrophys., 530}, 118.

\refs Pillai T., Wyrowski F. et al. (2006) {\it Astron. Astrophys., 450}, 569.



\refs Plambeck R. L., Bolatto A. D. et al. (2013) {\it Astrophys. J., 765}, 40.

\refs Plume R., Jaffe D., Evans N. et al. (1997) {\it Astrophys. J., 476}, 730.


\refs Preibisch T., Ratzka T. et al. (2011) {\it Astron. Astrophys., 530}, 40.

\refs Price D. J. and Monaghan J. J. (2004) {\it Mon. Not. R. Astron. Soc., 348}, 123.





\refs Qiu K., Zhang Q. et al. (2012) {\it Astrophys. J., 756}, 170.

\refs Qiu K., Zhang Q. et al. (2007) {\it Astrophys. J., 654}, 361.

\refs Qiu K., Zhang Q. et al. (2008) {\it Astrophys. J., 685}, 1005.

\refs Ragan S. E., Bergin E. A. et al. (2009) {\it Astrophys. J., 698,} 324.


\refs Ragan S. E. et al. (2012) {\it Astron. Astrophys., 547,} 49.

\refs Rathborne J. M. et al. (2006) {\it Astrophys. J., 641,} 389.



\refs Reid M. J., Argon A. L. et al. (1995) {\it Astrophys. J., 443}, 238.


\refs Robitaille T. P. et al. (2006) {\it Astrophys. J. Suppl., 167}, 256.

\refs Rod\'on J. A. et al. (2012) {\it Astron. Astrophys., 545,} 51.

\refs Roman-Duval J. et al. (2010) {\it Astrophys. J., 723}, 492.

\refs Rosen A. L., Krumholz M. R. et al. (2012) {\it Astrophys. J., 748}, 97.

\refs Sabbi E., Sirianni M., Nota A. et al. (2008) {\it Astron. J., 135}, 173.

\refs Sakai T., Sakai N., Furuya K. et al. (2012) {\it Astrophys. J., 742}, 140.

\refs Sakai T., Sakai N. et al. (2008) {\it Astrophys. J., 678}, 1049.

\refs San Jos{\'e}-Garc{\'{\i}}a I. et al. (2013) {\it Astron. Astrophys., 553}, 125.

\refs Sana H., de Mink S., de Koter A. et al. (2012) {\it Science, 337}, 444.

\refs S\'anchez-Monge \'A. et al. (2013a) {\it Astron. Astrophys., 552}, L10.

\refs S\'anchez-Monge \'A. et al. (2013b), {\it Astron. Astrophys., 552}, L10.


\refs S\'anchez-Monge \'A., Palau A., Fontani F. et al. (2013c) {\it Mon. Not. R. Astron. Soc., 432}, 3288.

\refs Sanhueza P., Jackson J. M. et al. (2013) {\it Astrophys. J., 773}, 123.

\refs Santangelo G., Testi L. et al. (2009) {\it Astron. Astrophys., 501}, 495.

\refs Schneider F. R. N., Izzard R. et al. (2014) {\it Astrophys. J., 780}, 117.

\refs Schneider N. et al. (2010) {\it Astron. Astrophys., 520}, 49.

\refs Selier R. et al. (2011) {\it Astron. Astrophys., 529}, 40.

\refs Sewilo M. et al. (2011) {\it Astrophys. J. Suppl., 194}, 44.



\refs Shu F. H. (1977) {\it Astrophys. J., 214}, 488.

\refs Shu F. H. et al. (1987) {\em Ann. Rev. Astron. Astrophys., 25}, 23.  


\refs Shu F. H., Najita J. R., Shang S., and Li Z.-Y. (2000) {\it Protostars and Planets IV}, (V. Mannings), Tucson, Univ. Arizona, 789.



\refs Smith R. J. et al. (2011) {\it Mon. Not. R. Astron. Soc., 411}, 1354.

\refs Smith R. J. et al. (2009) {\it Mon. Not. R. Astron. Soc., 400}, 1775.

\refs Smith R. J., Shetty R. et al. (2013) {\it Astrophys. J., 771}, 24.


\refs Sollins P., Hunter T., and Battat J. (2004) {\it Astrophys. J., 616}, L35.

\refs Sollins P., Zhang Q., Keto E. et al. (2005) {\it Astrophys. J., 624}, L49.

\refs Solomon P. M., Rivolo A. R. et al. (1987) {\it Astrophys. J., 319}, 730.


\refs Sridharan T. K. et al. (2014) {\it Astrophys. J., 783}, L31. 



\refs Su Y.-N., Liu S.-Y., Chen H. et al. (2012) {\it Astrophys. J., 744}, L26.

\refs Suttner G., Yorke H., and Lin D. (1999) {\it Astrophys. J., 524}, 857.

\refs Tan J. C. (2004) {\it Astrophys. J., 607}, L47.

\refs Tan J. C., Krumholz, M. R. et al. (2006) {\it Astrophys. J., 641}, L121.

\refs Tan J. C. and McKee C. F. (2003) astro-ph/0309139.

\refs Tan J. C. and McKee C. F. (2004) {\it Astrophys. J., 603}, 383.

\refs Tan J. C., Shaske S., and Van Loo S. (2013a) {\it IAU Symp., 292}, 19.

\refs Tan J. C., Kong S. et al.~(2013b) {\it Astrophys. J., 779}, 76.

\refs Tanaka K. E. I. and Nakamoto T. (2011) {\it Astrophys. J., 739}, L50.

\refs Tanaka K. E. I., Nakamoto T. et al. (2013) {\it Astrophys. J., 773}, 155.

\refs Tang Y-W. et al. (2009) {\it Astrophys. J., 700}, 251.


\refs Taquet V., Peters P.~S. et al. (2013) {\it Astron. Astrophys., 550}, 127.






\refs Testi L., Tan J. C. and Palla F. (2010) {\it Astron. Astrophys., 522}, 44.


\refs Torstensson K. J. E. et al. (2011) {\it Astron. Astrophys., 529}, 32.


\refs Turner J. L., and Beck S. C. (2004) {\it Astrophys. J., 602}, L85.


\refs Urban A., Martel H. and Evans N. (2010) {\it Astrophys. J., 710}, 1343.

\refs Vaidya B. et al. (2011) {\it Astrophys. J., 742}, 56.

\refs van der Tak F.~F.~S. et al. (2010) {\it Astron. Astrophys., 518}, L107.

\refs van der Wiel M.~H.~D. et al. (2013) {\it Astron. Astrophys., 553}, 11.





\refs Vasyunina T., Vasyunin A.~I. et al. (2012) {\it Astrophys. J., 751}, 105.

\refs Vink J. S., de Koter A. et al. (2001) {\it Astron. Astrophys., 369}, 574.


\refs Viti S. et al. (2004) {\it Mon. Not. R. Astron. Soc., 354}, 1141.


\refs Vlemmings W. et al. (2010) {\it Mon. Not. R. Astron. Soc., 404}, 134.


\refs Wang K.-S. et al. (2012) {\it Astron. Astrophys., 543}, 22.


\refs Wang P., Li Z., Abel T. et al. (2010) {\it Astrophys. J., 709}, 27.

\refs Wang Y., Zhang Q., Pillai, T. et al. (2008) {\it Astrophys. J., 672}, L33.


\refs Weidner C. et al. (2013) {\it Mon. Not. R. Astron. Soc., 434}, 84.

\refs Werner M. W., Capps R. W. et al. (1983) {\it Astrophys. J., 265}, L13.

\refs Whitehouse S. C. and Bate M. R. (2004) {\it Mon. Not. R. Astron. Soc., 353}, 1078.



\refs Williams J. P. and Garland C. A. (2002) {\it Astrophys. J., 568}, 259.

\refs Wolfire M. G. and Cassinelli J. (1987) {\it Astrophys. J., 319}, 850.

\refs Wood D. O. S. and Churchwell E. (1989) {\it Astrophys. J., 340}, 265.

\refs Wu J. and Evans N. J. (2003) {\it Astrophys. J., 592}, L79.

\refs Wu Y., Qin S.-L., Guan X. et al. (2009) {\it Astrophys. J., 697}, L116.


\refs Wu Y., Zhang Q., Chen H. et al. (2005a) {\it Astron. J., 129}, 330. 

\refs Wu Y., Zhu M., Wei Y. et al. (2005b) {\it Astrophys. J., 628}, L57.

\refs Wyrowski F. et al. (2012) {\it Astron. Astrophys., 542}, L15. 


\refs Y{\i}ld{\i}z U.~A. et al. (2013) {\it Astron. Astrophys.}, 556, 89. 

\refs Yorke H. and Sonnhalter C. (2002) {\it Astrophys. J., 569}, 846.

\refs Zapata L. A., Palau A., Ho P. et al. (2008) {\it Astrophys. J., 479}, L25.

\refs Zapata L. A. et al. (2013) {\it Astrophys. J., 765}, L29.

\refs Zhang Q., Sridharan T. K. et al. (2007) {\it Astrophys. J., 470}, 269.

\refs Zhang Q., Wang Y., Pillai T. et al. (2009) {\it Astrophys. J., 696,} 268.

\refs Zhang Y. and Tan J. C. (2011) {\it Astrophys. J., 733}, 55.

\refs Zhang Y., Tan J. C. et al. (2013a) {\it Astrophys. J., 766}, 86.
\refs Zhang Y., Tan J. C. et al. (2013b) {\it Astrophys. J., 767,} 58.

\refs Zhang Y., Tan J. C. et al. (2014) {\it Astrophys. J., 788,} 166.

\refs Zinnecker H. and Yorke H. (2007) {\em Ann. Rev. Astron. Astrophys. 45}, 481.

}
\end{document}